\documentclass[journal]{IEEEtran}
\usepackage{amsmath,amsfonts}
\usepackage{algorithm}
\usepackage{array}
\usepackage{textcomp}
\usepackage{stfloats}
\usepackage{url}
\usepackage{verbatim}
\usepackage{graphicx}
\usepackage{cite}
\usepackage{amsmath}
\usepackage{amssymb}
\usepackage{epstopdf}
\usepackage{subfigure}
\usepackage[section]{placeins}
\usepackage{graphicx}
\usepackage{makecell}
\usepackage{float}
\usepackage{booktabs}
\usepackage{picins}
\usepackage{xcolor}
\usepackage{algpseudocode}
\begin{document}

\title{ A Fast Robust Adaptive filter using Improved Data-Reuse Method}

\author{Yi Peng, Haiquan Zhao,~\IEEEmembership{Senior Member,~IEEE,} and Jinhui Hu 
\thanks{This work was partially supported by National Natural Science Foundation of China (grant: 62171388, 61871461, 61571374). Yi Peng, Haiquan Zhao,  and Jinhui hu  are with the Key Laboratory of Magnetic Suspension Technology and Maglev Vehicle, Ministry of Education, School of Electrical Engineering Southwest Jiaotong University Chengdu, China.  (e-mail: $hqzhao\_swjtu@126.com$; $pengyi1007@163.com$; $jhhu swjtu@126.com$)
 
Corresponding author: Haiquan Zhao.
}}



\maketitle

\begin{abstract}
Adaptive filter in complex scenarios demands algorithms that integrate fast convergence, low complexity, and robust performance under diverse noise conditions. To address this challenge, we propose a online censoring robust total generalized adaptive filter using improved data-reused method (RTGA-IDROC) algorithm.  The proposed RTGA variant possesses the advantages of both the total least squares (TLS) strategy and the robust generalized adaptive (RGA) function. This algorithm not only effectively handles input noise under the errors-in-variables (EIV) model but also achieves excellent performance across diverse noise environments. Furthermore, to meet the high demand for convergence speed in practical applications, an improved data reuse (IDR) method is introduced, enabling faster convergence in the early stages of iteration without compromising steady-state performance. The increased computational complexity brought by the IDR method is mitigated using the online censoring (OC) strategy. We also modify the OC threshold for real-valued algorithms, as the original threshold was defined for the complex domain. Beyond these algorithmic enhancements, a local stability analysis for the proposed algorithm is provided, and the theoretical steady-state mean-square deviation (MSD) is derived. Finally, simulation experiments in system identification and acoustic echo cancellation (AEC) scenarios validate the superior performance of the proposed algorithm.
\end{abstract}

\begin{IEEEkeywords}
errors-in-variables, data reuse, online censoring, generalized Gaussian noise, total least squares, acoustic echo cancellation.
\end{IEEEkeywords}

\section{Introduction}
\IEEEPARstart{W}{ITH} the advent of the least mean squre (LMS) algorithm \cite{LMS}, adaptive filtering techniques have found widespread applications in various fields, such as acoustic echo cancellation (AEC), system identification, and power system frequency estimation \cite{lv2023TSP,Ni2023,zayyani2020,ACTLS2025augmented,c1,c2,c3,c4,c5,c6,c7,c8,c11,c12}. However, the LMS algorithm and its variants primarily account for the effect of output noise during operation, while in practical applications, input signals are often also contaminated by noise due to human errors, sensor imperfections, and other issues. Such scenarios, where both input and output noises are present, are collectively described by the errors-in-variables (EIV) model \cite{EIV}. Due to their inability to effectively handle the impact of input noise, LMS-based algorithms tend to produce biased estimates when the input signal contains noise .

To address this issue, the gradient descent TLS (GDTLS) algorithm was proposed in \cite{GDTLS}. It effectively handles the EIV scenario where noise is present in both input and output signals. While GDTLS maintains strong performance under Gaussian noise, its performance severely degrades in the presence of impulsive noise in the output signal. This is because its cost function relies entirely on the second-order moment of the error. To mitigate this limitation, several algorithms have been developed using robust cost functions that are resistant to impulsive disturbances. Notable examples include the total
least mean M-estimate (TLMM) algorithm \cite{TLMM} derived from M-estimation, the total maximum versoria Criterion (TMVC) algorithm \cite{CTMVC} built on the MVC function, the maximum total correntropy (MTC) algorithm \cite{MTC,MTCC} based on correntropy theory,   and the logarithmic total least
square (LTLS) algorithm \cite{LTLS} utilizing a logarithmic cost function. Owing to the robustness of these cost functions against impulsive noise, all these methods achieve noticeable performance improvement over GDTLS in EIV models contaminated by impulsive noise.

However, real-world noise conditions are often more complex and are not limited to impulsive noise. When the noise environment contains generalized Gaussian noise, the performance of the aforementioned algorithms degrades significantly. To address this issue, the maximum total generalized correntropy (MTGC) algorithm \cite{MTGC} replaces the Gaussian kernel in maximum total correntropy (MTC) with a generalized Gaussian kernel \cite{GMCC}, adapting the order of the error in the cost function to better handle generalized Gaussian noise. Building on this, to further improve the steady-state performance under such noise, the robust generalized maximum Blake-Zisserman total correntropy (GMBZTC) and improved robust total logistic distance metrics (TACLDM) algorithms were proposed in \cite{GMBZTC} and \cite{TACLDM}, respectively. It can be observed that researchers have developed various algorithms based on different cost functions to cope with different types of noise, yet each algorithm only maintains optimal performance in specific noise environments. For instance, while MTGC outperforms GDTLS under impulsive or generalized Gaussian noise due to its robustness against these disturbances, it underperforms compared to GDTLS in a pure Gaussian noise environment. Notably, the robust generalized adaptive (RGA) algorithm with a flexible cost function was proposed in \cite{RGA1,RGA}. Its form can be flexibly altered by tuning parameters, enabling it to maintain optimal performance under different noise environments.

Unfortunately, the RGA algorithm is still developed based on the LMS framework and thus cannot effectively address the estimation bias caused by input noise. Therefore, it is necessary to propose a variant capable of handling input noise. Additionally, in practical applications, convergence speed has always been a key performance metric for algorithms. Data reuse is a common technique to improve convergence speed. However, most existing data reuse methods still primarily rely on repeatedly reusing the current data \cite{Data1,Data2,Data3,Data4}. This not only reduces the stability of the algorithm but also caps its performance ceiling to that of its normalized variant—for instance, the performance of data-reuse LMS (DRLMS) is bounded by that of normalized LMS (NLMS) \cite{1989anly}. It should also be noted that NLMS does not necessarily outperform LMS in all scenarios \cite{LMS_better}, which implies that existing DRLMS algorithms are not always superior to LMS. Moreover, data reuse often leads to a marked increase in computational complexity, which grows proportionally with the number of reuses.

To address all the aforementioned issues, this paper proposes the online censoring robust total generalized adaptive filter using improved data-reused method (RTGA-IDROC) algorithm, which employs the TLS strategy to mitigate the impact of input noise.  By introducing an improved data reuse (IDR) method, the algorithm not only accelerates convergence in the early stages of iteration but also achieves steady-state performance comparable to that of the original algorithm without data reuse.  Moreover, the increase in computational complexity is effectively constrained via an online censoring (OC) strategy \cite{OC}. It is worth emphasizing that existing OC strategies in adaptive filtering are predominantly designed for complex-domain applications \cite{OC1,OC2,OC33,OC4}. When applied in real-valued scenarios, however, the distribution used for threshold determination shifts from Rayleigh to half-normal, rendering conventional thresholds ineffective. Therefore, this paper introduces a new OC threshold tailored for the real domain. Extensive simulations finally validate the superior performance of the proposed algorithm in both system identification and  AEC scenarios. The main contributions of this paper are summarized as follows:

(1) The RTGA-IDROC algorithm is proposed, which achieves better performance than other competing algorithms in EIV models contaminated by generalized Gaussian noise.

(2) An IDR strategy is developed, enabling algorithms that adopt it to achieve faster convergence while maintaining steady-state performance on par with their original versions without IDR. In addition, a novel OC threshold suitable for real-domain algorithms is introduced and verified through simulations.

(3) A detailed analysis of the steady-state performance of the RTGA-OC algorithm is provided, accompanied by corresponding simulations.

(4) Extensive simulations are conducted to demonstrate the effectiveness and superior performance of the proposed algorithm in AEC application.

The remainder of this paper is organized as follows. Section II reviews the EIV model and presents the IDR method and real-domain OC strategy. Section III provides a detailed derivation of the RTGA-OC algorithm. Section IV presents a comprehensive analysis of the RTGA-OC algorithm, covering its local performance, mean stability, and steady-state behavior. The computational complexity of the RTGA-IDROC algorithm is analyzed in Section V. Section VI validates the algorithm's excellent performance and the theoretical findings through computer simulations. Finally, Section VII concludes the paper.
 
\begin{figure}[t]
	\centering
	\includegraphics[scale=0.19]{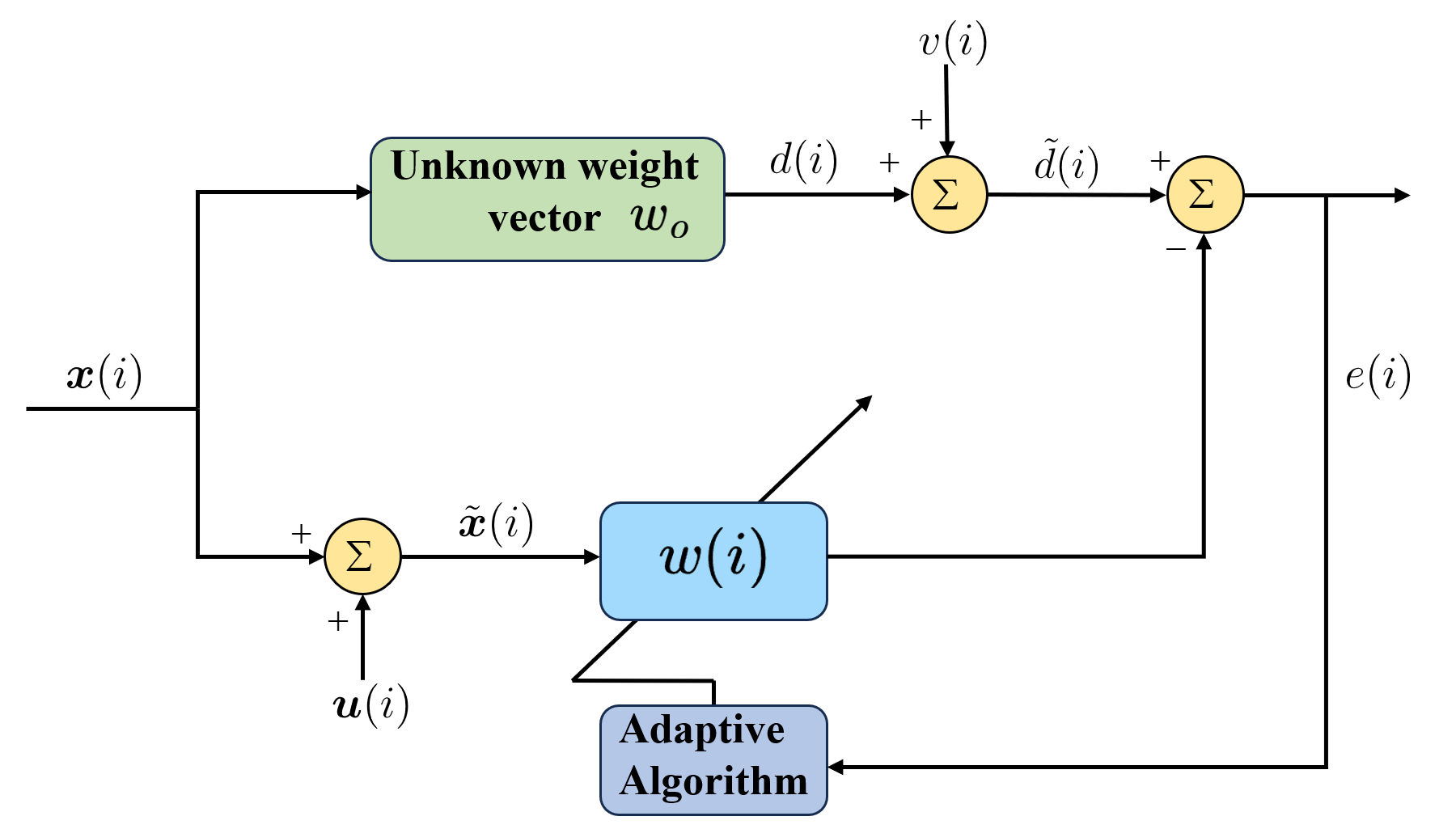} 
	\caption{EIV model}
\end{figure}

\section{System Model and Preliminaries}
\subsection{ EIV model}
As illustrated in Fig. 1, a linear model can be defined as $d(i)=\boldsymbol w_o^T \boldsymbol x(i)$, where $i$ is discrete time, $d(i)$  represents  desired signal, $\boldsymbol{x}(i)\in \mathbb{R}^{L\times1}$ and $\boldsymbol{w}_o \in \mathbb{R}^{L\times1}$ are the input vector and unknown weight vector.

Unlike traditional system identification models that only consider noise at the output, the EIV model considers the situation where both input and output vectors are affected by noise \cite{EIV}. This can be formulated as
\begin{equation}
    \tilde {\boldsymbol{x}}(i)=\boldsymbol{x}(i)+\boldsymbol{u}(i)
\end{equation}
\begin{equation}
        \tilde d(i) = d(i)+v(i)
\end{equation}
where $v(i)$ represents the output noise with variance $\sigma_o^2$, and $\boldsymbol{u}(i) \in \mathbb{R}^{L \times 1} $ denotes the input noise with variance $\sigma_i^2$, which is assumed to be independently and identically distributed (i.i.d). The error signal $e(i)$ and the actual signal $y(i)$ can be represented as
\begin{equation}
    e(i)=\tilde{d}(i)-y(i)
\end{equation}
\begin{equation}
    y(i)=\boldsymbol{w}(i)^T\tilde{\boldsymbol{x}}(i)
\end{equation}
where $\boldsymbol w (i) \in \mathbb{R}^{L \times 1}$

\subsection{An Improved Data Reuse method}
The DRLMS algorithm was first proposed in \cite{DR_orgin}. Its core principle involves reusing the current data sample multiple times to achieve a higher convergence rate. Numerous algorithms incorporating this data-reuse strategy have demonstrated performance improvement \cite{DR11,DR33,DR44}. However, since reusing the current sample does not alter the direction of weight vector update, the extent of convergence acceleration is limited. The performance ceiling is bounded by the NLMS algorithm \cite{1989anly}, and the method can also lead to degraded stability.

Subsequently, researchers proposed a new data reusing method called Unnormalized new DRLMS (UNDRLMS) \cite{UNDR}. This method employs the past $L$ input vectors to update the algorithm, thereby avoiding the use of repeated data pairs. Although this approach introduces new update directions and partially addresses the issues of traditional DRLMS, the reused data pairs remain temporally close to the current data pair. This proximity maintains high correlation among the reused input vectors. Geometrically, this corresponds to a small angle between the solution hyperplanes. Consequently, in most cases, the performance improvement remains limited.

To address these challenges, we propose an improved data reusing (IDR) method  that uniformly divides past data into $L_\text{reused}$ segments and selects points from different segments. This reduces the correlation among the reused input vectors, further enhancing the performance gains from the data reusing approach. The proposed method exhibits an initial convergence behavior similar to that of conventional DR method. During the steady state, however, the large intervals between the selected points allow the input vectors to be considered uncorrelated, enabling our method to achieve the same steady-state performance as algorithms without data reusing. The proposed data reuse method is summarized in Algorithm 1, where $L$ is filter order, $L_{\text{reuse}}$ denotes the number of data reuse times, $\lfloor \cdot \rfloor$ represents the floor function, and $g_{\text{cost}}$ indicates the gradient of the utilized cost function.
\begin{algorithm}[t]
	\caption{Weight Update with Reuse}
	\begin{algorithmic}[1]
		\State \textbf{Initialization}: $\boldsymbol{w}(0)=\textbf{0}$
	\For {$i > L$}
	    \State $\boldsymbol{w}_\text{temp}=\boldsymbol{w}(i-1)$
		\For{$ii = 1$ : $L_{\text{reuse}}$}
		\State $\text{idx} = L + \left\lfloor \dfrac{(i - L) \times ii}{L_{\text{reuse}} + 1} \right\rfloor$
		\State $e_{\text{temp}} = d(\text{idx}) -\boldsymbol{w}_{\text{temp}}^T \boldsymbol{x}(\text{idx}) $
		\State $\boldsymbol{w}_{\text{temp}} = \boldsymbol{w}_{\text{temp}} - \mu \boldsymbol{g}_{\text{cost}}$
		\EndFor
		\State $\boldsymbol{w}(i)=\boldsymbol{w}_\text{temp}$
		\State $e(i) = d(i) - \boldsymbol{w}(i)^T\boldsymbol{x}(i) $
		\State $\boldsymbol{w}(i+1)=\boldsymbol{w}(i)-\mu \boldsymbol{g}_\text{cost}$
	\EndFor	
	\end{algorithmic}
\end{algorithm}

Fig. 2 provides a graphical comparison of various DR methods, where \( s_i \), \( s_{i-1} \), and \( s_{idx} \) denote the solution hyperplanes for the input--output pairs \( \{d(i), \boldsymbol{x}(i)\} \), \( \{d(i-1), \boldsymbol{x}(i-1)\} \), and \( \{d(idx), \boldsymbol{x}(idx)\} \), respectively, under the simplifying assumption \( d(i) = \boldsymbol{w}_o^T \boldsymbol{x}(i) \). This figure integrates our proposed method with LMS for direct comparison against related LMS-type algorithms. To maintain consistency across the illustrations, every iteration of all the algorithms in the demonstration follows a unified execution flow: one update step followed by one reuse step.

As shown in Figs. 2(a), (b) and (c), the UNDR-LMS algorithm converges faster than traditional DR-LMS by incorporating past data pairs. However, its reuse of adjacent, highly correlated input vectors (resulting in a small \( \theta_s \)) limits the performance gain. Our proposed method overcomes this by selecting less correlated input vectors, which increases \( \theta_s \) and, as visually evident, enables a more rapid approach to the optimal weight vector with each update.

\begin{figure}[htbp]
	\centering
	\begin{minipage}{0.24\textwidth}
		\centering
		\includegraphics[width=\linewidth]{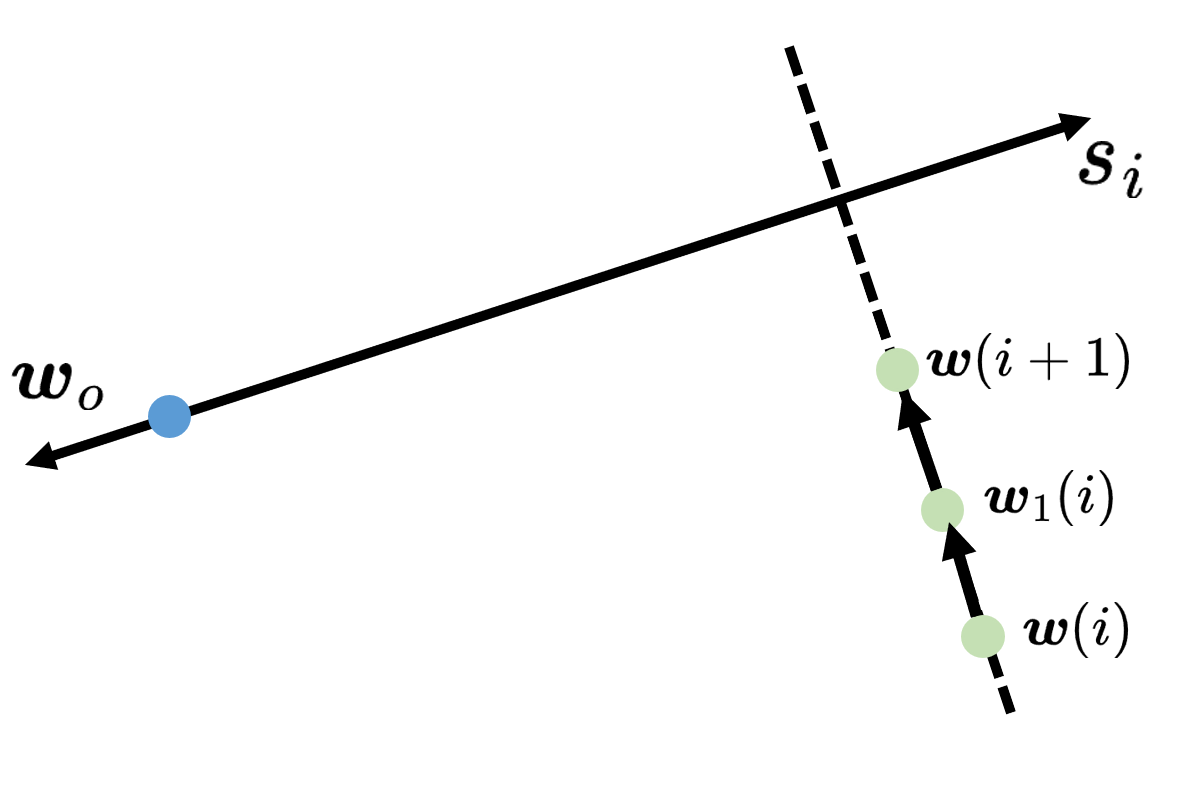}
		\\\text{(a)} $L_\text{reused} =1 $
		\label{fig:sub1}
	\end{minipage}
	\hfill
	\begin{minipage}{0.24\textwidth}
		\centering
		\includegraphics[width=\linewidth]{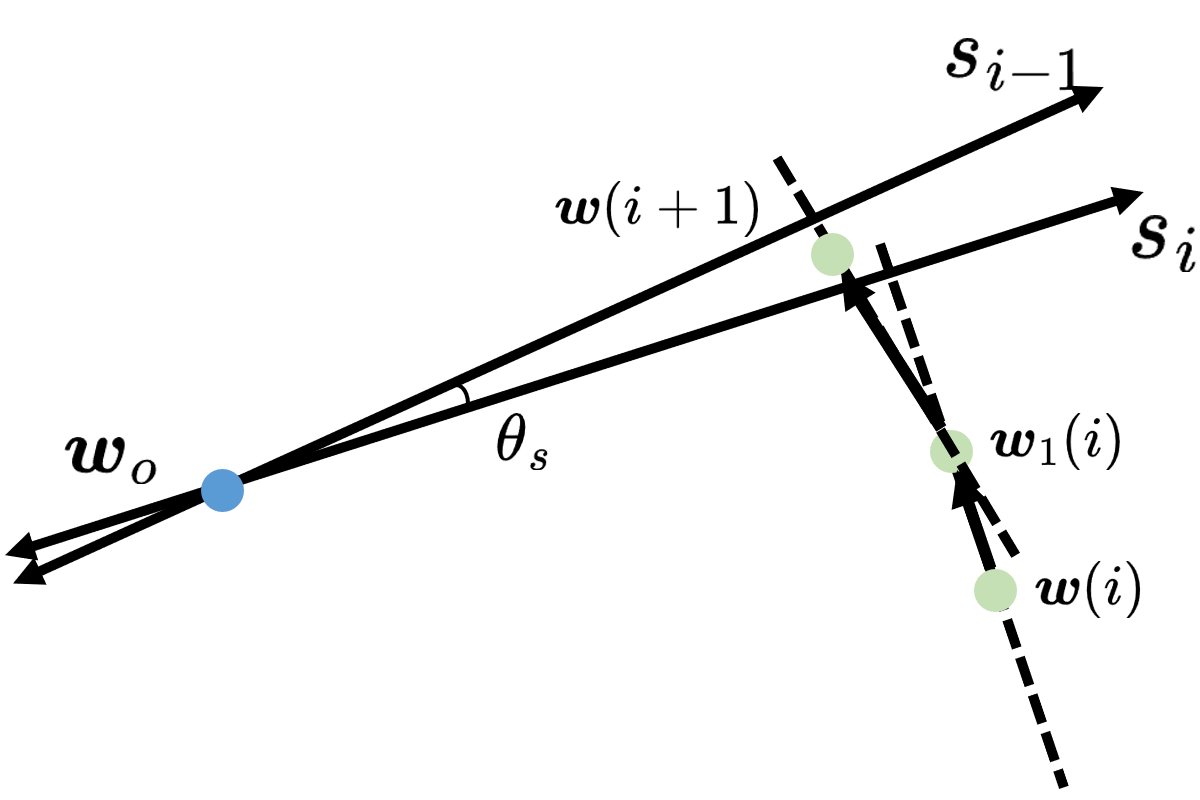}
        \\\text{(b)}$L_\text{reused} =1 $
		\label{fig:sub2}
	\end{minipage}
	
	\vspace{0.5cm}
	
	\begin{minipage}{0.25\textwidth}
		\centering
		\includegraphics[width=\linewidth]{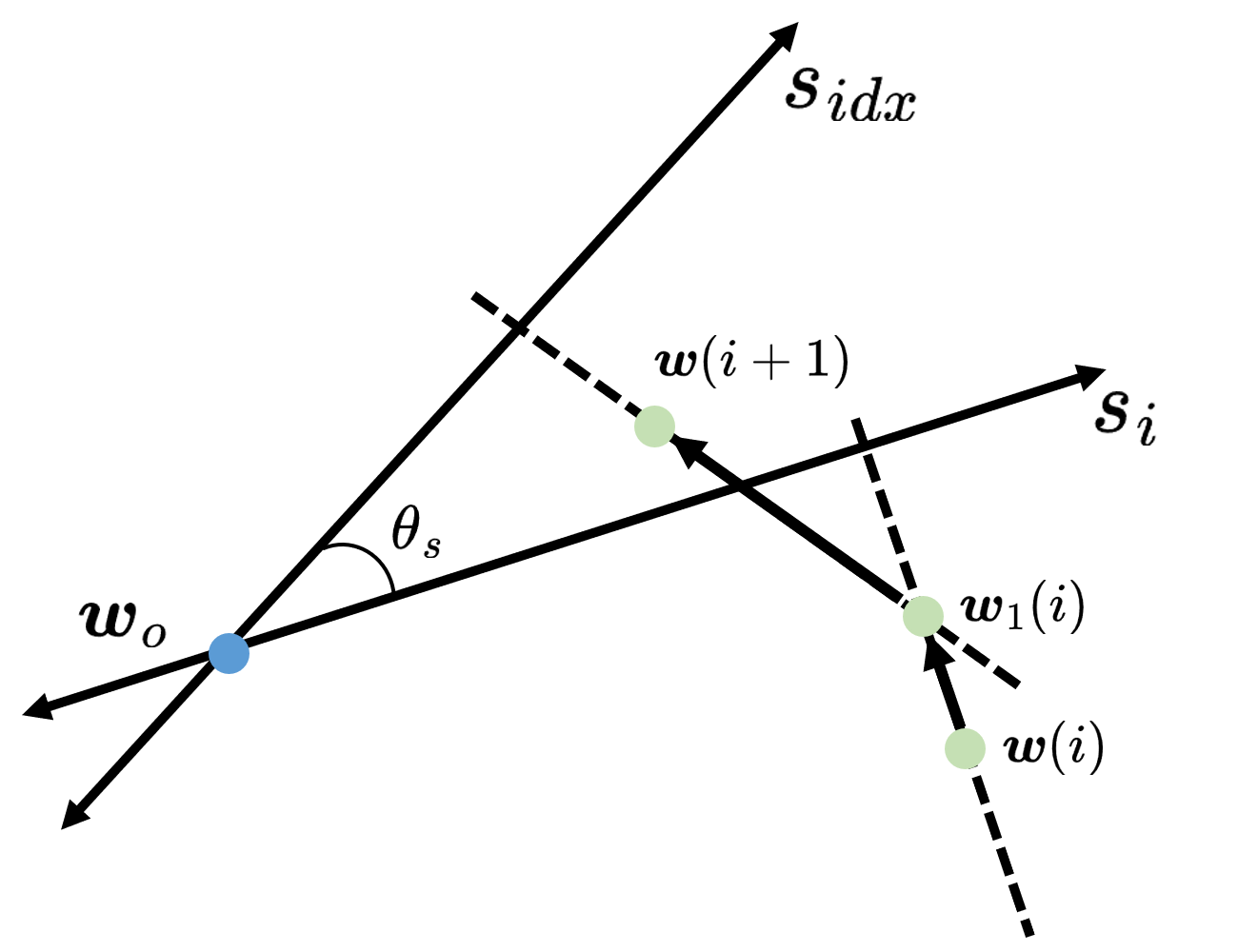}
		\\\text{(c)}$L_\text{reused} =1 $
		\label{fig:sub3}
	\end{minipage}
	\caption{Graphical depiction of the underlying principle for the (a) DR-LMS, (b) UNDR-LMS, and (c) IDR-LMS algorithm.}
	\label{fig:three_images}
\end{figure}

\subsection{The OC method for real-valued algorithm}
Online censoring is a computational complexity reduction technique \cite{OC} that selectively eliminates less informative observations and updates only statistically significant ones. The corresponding selection strategy can be formulated as
\begin{equation}
	(\mathcal{Z}(i),\mathcal{C}(i)):=\begin{cases}
		(\tilde{d}(i),0),& \text{if} |e(i)|\geq \kappa \sigma_o\\ 
		(\circ ,1),& \text{otherwise}
	\end{cases}
\end{equation}
where $\kappa$, $\circ$, and $\mathcal{C}(i)$ $> 0$, are the censoring threshold, the
censored real-valued data stream at any time index $i$, and the binary censoring variable, respectively. In (3), if $\mathcal{C}(i) = 1$, the data point $d(i)$ is considered less informative and is discarded without processing. If $\mathcal{C}(i) = 0$,  $d(i)$ is identified as highly informative and is retained for immediate use. It is noteworthy that existing OC-based adaptive filtering algorithms have been developed solely for the complex domain. However, when the OC strategy is applied to real-domain algorithms, the original threshold calculation formula becomes inapplicable. This is because the distribution of $|e(i)|$ shifts from Rayleigh \cite{OC1} to half-normal. After some mathematical operations, the threshold calculation formula for real-domain OC algorithms is given as
\begin{equation}\label{6}
	\kappa=\sqrt{2}{erf^{-1}(P_{ce})}
\end{equation}
where $erf(x)=2\int_{0}^{x}e^{-t^2}dt/\sqrt{\pi}$, target censoring ratio $P_{ce} = \frac{I-p}{I}$, $I$ is the total number of data samples and $p$ is the number of informative data. 
The validity of the proposed threshold will be verified in the Section VI-A by comparing the preset censoring ratio $P_{ce}$ with its estimated value $\hat{P}_{ce}$.

\section{The RTGA-IDROC algorithm}
In this paper, we propose the RTGA-IDROC algorithm, which enhances robustness to input noise by introducing the TLS method. This algorithm can not only effectively suppress input noise but also, due to the flexibility of its functional form, exhibits good adaptability in different output noise environments. 

Furthermore, by leveraging the advantages of both the IDR method and the OC strategy, the RTGA-IDROC algorithm achieves accelerated convergence without compromising steady-state performance, while effectively restraining the growth in computational complexity.
It is worth noting that we introduce a real-valued OC threshold mechanism, as designed in the previous section, to evaluate each iteration: weight updates are executed only when the input data contains sufficient information. This evaluation mechanism is applied to both conventional updates and the data reuse process, thereby effectively controlling computational cost without compromising convergence speed. 

Finally, the cost function of the proposed RTGA-IDROC algorithm can be expressed as
\begin{equation}
	J(i)=\begin{cases}
		E(\frac{|a-b|}{a}\left [(\frac{c|\frac{e(i)}{\left \| \bar {\boldsymbol w}(i)\right \|}|^b}{|a-b|}+1)^\frac{a}{b}-1\right]),& if |e(i)|\geq \kappa \sigma_e\\ 
		0,& \text{otherwise}
	\end{cases}
\end{equation}
where $E(\cdot)$ is denotes the expectation operator,  $a \in \mathbb{R}$, $b>0$ are shape parameters, $\bar {\boldsymbol w}(i)\triangleq [\sqrt{\phi}, -\boldsymbol{w}^T ]^T$, where $\phi=\frac{\sigma_o^2}{\sigma_i^2}$, and $c$ is scale parameter. Considering the influence of impulsive noise, continued use of the traditional formula to estimate $\sigma_e$ would lead to a significant overestimation. Accordingly, the following estimation formula is adopted
\begin{equation}\label{8}
	\sigma_e(i+1)=\tau 	\sigma_e(i) + 1.483(1-\tau)\rho\{|e(i)|\}
\end{equation}
where $\rho\{|e(i)|\}=\text{med}\{|e(i)|,\cdots,|e(i-Nw+1)|\}$, $\text{med}(\cdot)$ is median
operation, 1.483 is correction factor, and $N_w$ is  the length of the sliding window, typically between 7 and 15 \cite{OC3}.

 The RTGA cost function is evidently undefined at certain points. Interestingly, we observe that the limit forms of the cost function at these points are identical in form to those of several common TLS-based adaptive filtering algorithms. Therefore, we provide the analysis of these distinct points as follows
\begin{itemize}
	\item At $a=b$ \begin{equation}\label{a=b}
			\lim_{a\to b}J(i)=\begin{cases}
			\frac{c}{b}|\frac{e(i)}{\left \| \bar {\boldsymbol w}(i)\right \|}|^b,& if |e(i)|\geq \kappa \sigma_e\\ 
			0,& \text{otherwise}
		\end{cases}
	\end{equation}
	\item At $a=0$ \begin{equation}\label{a=0}
		\lim_{a\to b}J(i)=\begin{cases}
			\text{log}(\frac{c}{b}|\frac{e(i)}{\left \| \bar {\boldsymbol w}(i)\right \|}|^b+1),& if |e(i)|\geq \kappa \sigma_e\\ 
			0,& \text{otherwise}
		\end{cases}
	\end{equation}
	\item At $a \to -\infty$ \begin{equation}\label{a=-inf}
	\lim_{a\to b}J(i)=\begin{cases}
		1-\exp(-\frac{c}{b}|\frac{e(i)}{\left \| \bar {\boldsymbol w}(i)\right \|}|^b),& if |e(i)|\geq \kappa \sigma_e\\ 
		0,& \text{otherwise}
	\end{cases}
\end{equation}
\end{itemize}

\textbf{Remark} \textbf{1}: When $a$ approaches $b$, Eq. (\ref{a=b}) reduces to the total least mean p-norm (TLMP) algorithm \cite{TLMP} . Specifically, it corresponds to the total least mean fourth (TLMF) algorithm \cite{TLMF} when $b=4$, and simplifies to the classical TLS algorithm when $b=2$. When $a \to 0$ and $b=2$, Eq. (\ref{a=0}) becomes equivalent to the LTLS algorithm. As $a \to -\infty$, Eq. (\ref{a=-inf}) converges to the MTGC algorithm, particularly simplifying to the MTC algorithm when $b=2$. This flexible functional form demonstrates its strong adaptability to diverse noise environments. Furthermore, the proposed algorithm fails to converge when $a \to +\infty$, and thus this case is excluded from subsequent analysis.

The corresponding gradient vector can be derived as
\begin{equation}\label{g0}
	\begin{aligned}
     &	\boldsymbol g_J(i)=\frac{\partial J(i)}{\partial 	\boldsymbol w(i)}=\begin{cases}
     	\boldsymbol g(i),& if |e(i)|\geq \kappa \sigma_e\\ 
     	0,& \text{otherwise}
     \end{cases}
    \end{aligned}
\end{equation}
where 
\begin{equation}
	\boldsymbol g(i)=E(-c(\frac{c|\frac{e(i)}{\left \| \bar {\boldsymbol w}(i)\right \|}|^b}{|a-b|}+1)^\frac{a-b}{b}{|\frac{e(i)}{\left \| \bar {\boldsymbol w}(i)\right \|}|^{b-2}}\psi (i))
\end{equation}
\begin{equation}
	\psi(i)= \frac{\tilde {\boldsymbol{x}}(i)e(i)+\frac{e^2(i)\boldsymbol{w}(i)}{\|\bar {\boldsymbol w}(i)\|^2}}{\|\bar {\boldsymbol w}(i)\|^2}
\end{equation}

Then, we can further obtain the instantaneous gradient vector as
\begin{equation}
	\begin{aligned}\label{g1}
		&	\hat {\boldsymbol g}_J(i)=\frac{\partial J(i)}{\partial 	\boldsymbol w(i)}=\begin{cases}
		\hat {\boldsymbol g}(i),& if |e(i)|\geq \kappa \sigma_e\\ 
			0,& \text{otherwise}
		\end{cases}
	\end{aligned}
\end{equation}
where
\begin{equation}
		\hat {\boldsymbol g}(i)= 	-c(\frac{c|\frac{e(i)}{\left \| \bar {\boldsymbol w}(i)\right \|}|^b}{|a-b|}+1)^\frac{a-b}{b}{|\frac{e(i)}{\left \| \bar {\boldsymbol w}(i)\right \|}|^{b-2}}\psi (i)
\end{equation}

Using the gradient descent method, we can get weight vector update formula as
\begin{equation}\label{update}
	\boldsymbol{w}(i+1)=\boldsymbol{w}(i)-\mu \hat {\boldsymbol{g}}(i)
\end{equation}

Finally, the update rule (\ref{update})  is integrated with the IDR method, yielding the complete RTGA-IDROC algorithm. The detailed iterative procedure is summarized in Algorithm 2.

\begin{algorithm}[t]
	\caption{RTGA-IDROC}
	\begin{algorithmic}[1]
		\State \textbf{Parameters}: $L_\text{reused}$, $\mu$, $a$, $b$, $c$
		\State \textbf{Initialization}: $\boldsymbol{w}(0)=\textbf{0}$
		\For {$i > L$}
		\State $\boldsymbol{w}_\text{temp}=\boldsymbol{w}(i-1)$
		\For{$ii = 1$ : $L_{\text{reuse}}$}
		\State $\text{idx} = L + \left\lfloor \dfrac{(i - L) \times ii}{L_{\text{reuse}} + 1} \right\rfloor$
		\State $e_{\text{temp}} = d(\text{idx}) -\boldsymbol{w}_{\text{temp}}^T \tilde{\boldsymbol{x}}(\text{idx}) $
		\State \textbf{The instantaneous gradient $\hat{\boldsymbol g}(i)$ can be obtained via (\ref{g1})}
		\State $\boldsymbol{w}_{\text{temp}} = \boldsymbol{w}_{\text{temp}} - \mu \hat{\boldsymbol g}(i)$
		\EndFor
		\State $\boldsymbol{w}(i)=\boldsymbol{w}_\text{temp}$
		\State $e(i) = d(i) - \boldsymbol{w}(i)^T \tilde{\boldsymbol{x}}(i) $
		\State \textbf{The instantaneous gradient $\hat{\boldsymbol g}(i)$ can be obtained via (\ref{g1})}
		\State $\boldsymbol{w}(i+1)=\boldsymbol{w}(i)-\mu \hat{\boldsymbol g}(i)$
		\EndFor	
	\end{algorithmic}
\end{algorithm}

\begin{figure}[htbp]
	\centering
	\begin{minipage}{0.24\textwidth}
		\centering
		\includegraphics[width=\linewidth]{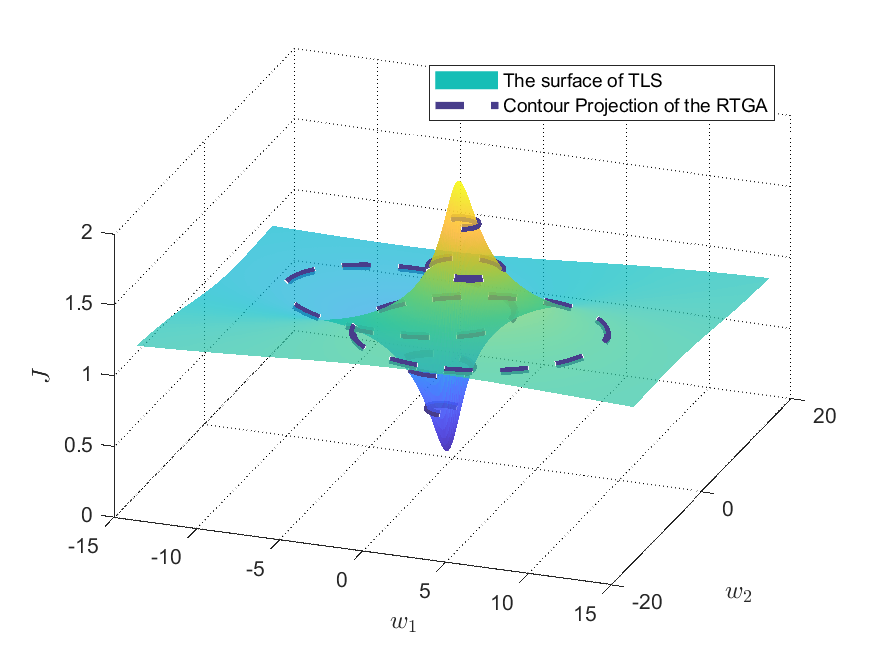}
		\\\text{(a)}
	\end{minipage}
	\hfill
	\begin{minipage}{0.24\textwidth}
		\centering
		\includegraphics[width=\linewidth]{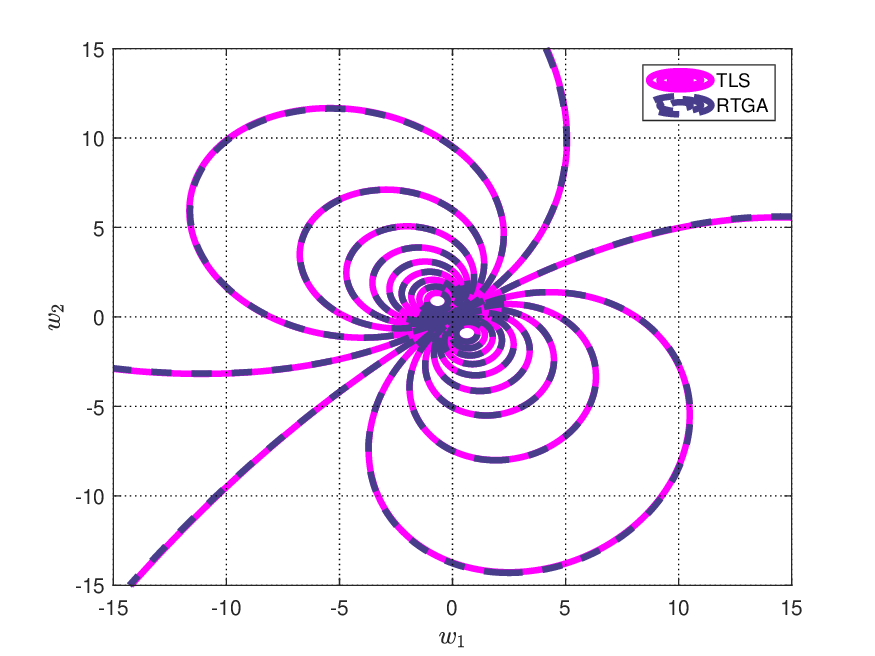}
		\\\text{(b)}
	\end{minipage}
	
	\begin{minipage}{0.24\textwidth}
		\centering
		\includegraphics[width=\linewidth]{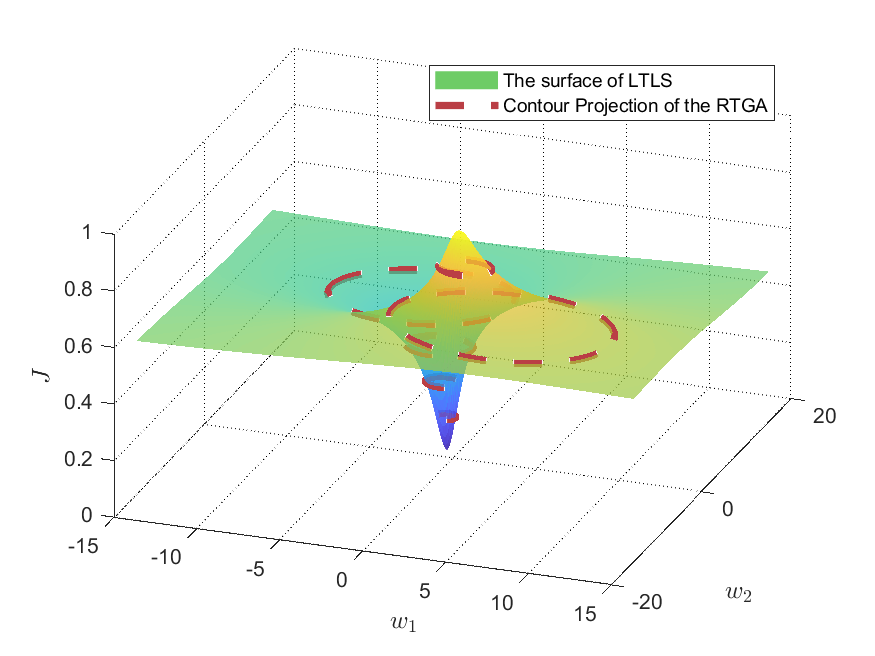}
		\\\text{(c)}
	\end{minipage}
	\hfill
	\begin{minipage}{0.24\textwidth}
		\centering
		\includegraphics[width=\linewidth]{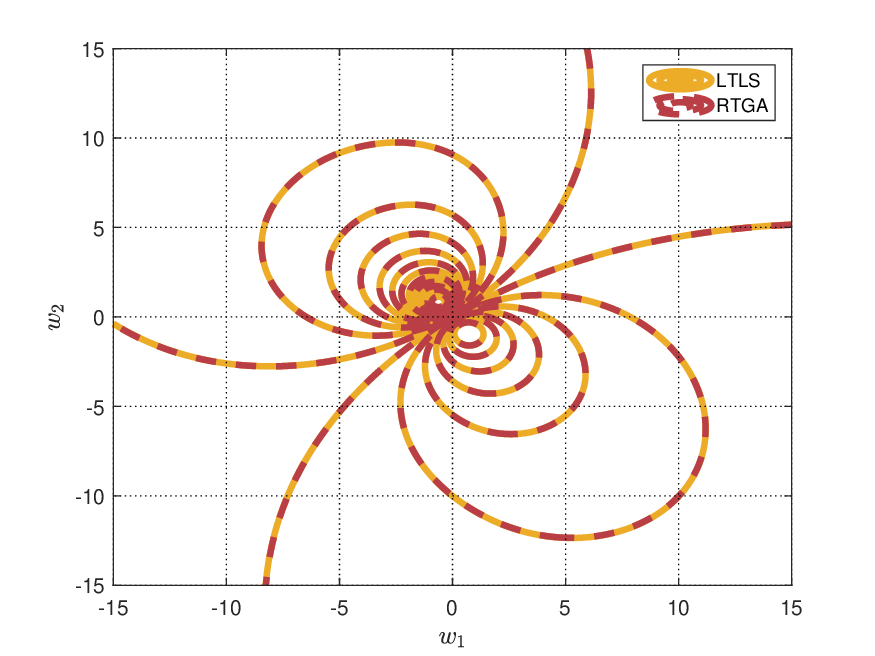}
		\\\text{(d)}
	\end{minipage}
	
	\begin{minipage}{0.24\textwidth}
		\centering
		\includegraphics[width=\linewidth]{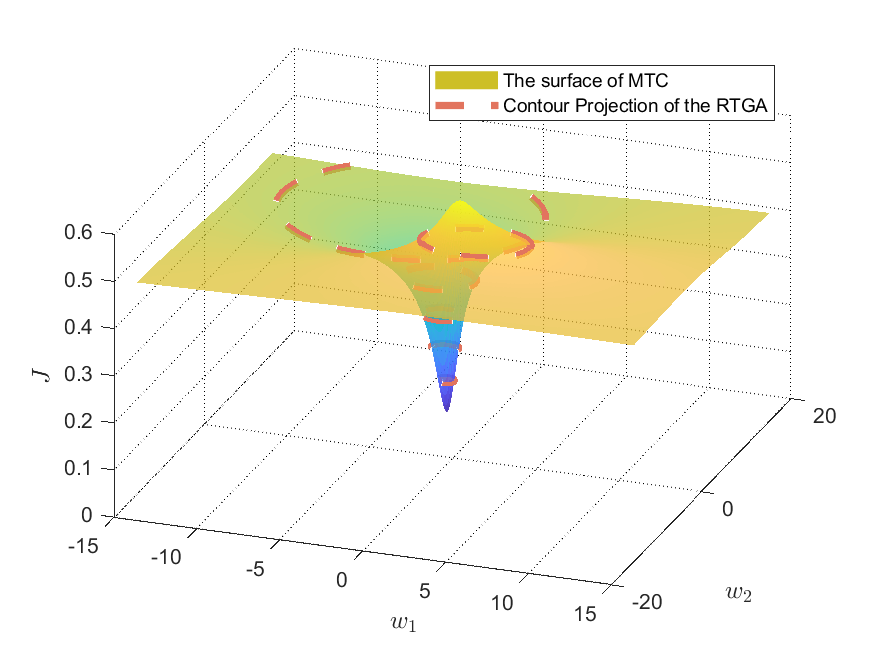}
		\\\text{(e)}
	\end{minipage}
	\hfill
	\begin{minipage}{0.24\textwidth}
		\centering
		\includegraphics[width=\linewidth]{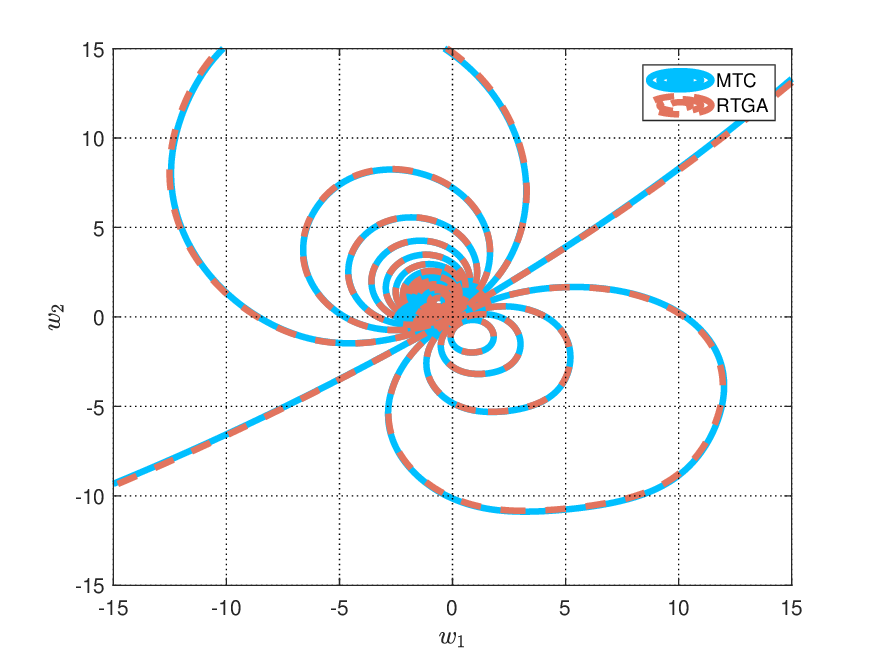}
		\\\text{(f)}
	\end{minipage}
	
	\caption{Comparative visualization of the RTGA function against classical TLS-type cost functions}
	\label{fig:six_subfigures}
\end{figure}

\textbf{Remark} \textbf{2}: To verify the conclusions in \text{Remark 1}, Fig. 3 projects the contour lines of the RTGA function onto the 3D surfaces of several common TLS-type cost functions and compares the RTGA contours with these TLS-type cost functions. All simulations use the following unified experimental settings: $\sigma_o^2=\sigma_i^2=0.1$, filter order $L=2$, $\textbf{R}=\mathbf{I}$, and $\boldsymbol{w}_o=[-0.6,0.8]^T$, where the matrix $\textbf{R}$ (defined in A2 of Section IV) is the identity matrix. The RTGA parameters are configured as follows across subfigures: (a) and (b) with $a=1.999$, $b=2$, $c=2$; (c) and (d) with $a=0.001$, $b=2$, $c=2$; (e) and (f) with $a=-1000$, $b=2$, $c=0.02$. As shown in Fig. 3, under the three different cases corresponding to (8), (9), and (10), the proposed RTGA cost function closely approximates the respective TLS-based cost functions in all scenarios. It is precisely this structural flexibility that enables the RTGA algorithm to achieve strong adaptability to various noise environments.

\textbf{Remark} \textbf{3}: It is noteworthy that the proposed RTGA-OC method does not employ the second stage threshold judgment used in the ROC method of \cite{OC,OC1}. This is because the second stage threshold only becomes effective when dealing with large amplitude impulsive noise, which occurs with relatively low probability in practice. For instance, in most algorithms \cite{TLMM,MTC,GMCC,liu2020robust} that consider impulsive noise as outliers, the occurrence probability is typically around $1\%$. Futhermore, the second stage threshold not only introduces significant computational overhead but also fails to provide robustness against other types of outliers such as low amplitude generalized Gaussian noise, unlike the RTGA function. Although removing the second stage threshold may reduce the estimated OC probability $\hat{P}_{ce}$ 
 , the RTGA algorithm inherently maintains strong robustness against impulsive noise. Thus, eliminating this threshold not only reduces computational waste but also preserves the algorithm's performance.

\section{Performance Analysis}
Our performance analysis focuses primarily on the core RTGA-OC algorithm. This approach is justified by the observation in Section II-B that the IDR method exhibits initial convergence behavior identical to traditional DR methods (which have been extensively analyzed in the literature), while achieving steady-state performance equivalent to algorithms operating without IDR. To streamline the analytical process, we adopt several commonly used assumptions\cite{MTC,MTGC,GMBZTC}.

A1: Both the output noise $v(i)$ and input noise $\boldsymbol{u}(i)$ follow the generalized Gaussian distribution (GGD), defined by $f(x;\alpha,\beta)$ \cite{GMBZTC}.

A2: The matrix $\textbf{R}=E[\boldsymbol{x}(i)\boldsymbol{x}(i)^T]$ is positive definite and full rank.

A3: The output noise $v(i)$, the input noise $\boldsymbol{u}(i)$ and input signal $\boldsymbol{x}(i)$ are independent of each other.

\subsection{Local Extreme Point}
The introduction of the OC strategy prevents continuous updates of the filter weights in (\ref{update}), as the gradient is occasionally set to zero, thereby complicating the performance analysis. Therefore, by introducing a update probability $P_t$, (\ref{g0}) can be rewritten as
\begin{equation}\label{g3}
	\boldsymbol{g}_t(i)=P_tE(-c(\frac{c|\frac{e(i)}{\left \| \bar {\boldsymbol w}(i)\right \|}|^b}{|a-b|}+1)^\frac{a-b}{b}{|\frac{e(i)}{\left \| \bar {\boldsymbol w}(i)\right \|}|^{b-2}}\psi (i))
\end{equation}
where $P_t=Pr\{{|e(i)|\geq \kappa \sigma_v}\}=1-P_{ce}$, $Pr(\cdot) $ denotes the probability operator.

As observed in Fig. 3, both RTGA and other TLS-type algorithms exhibit two critical points. Since we employ gradient descent for algorithm updates, the convergent point should be a local minimum. Therefore, to prove that the obtained weight vector indeed corresponds to a local minimum, detailed analysis of the extreme points is necessary.

It should first be proven that (\ref{g3}) equals zero at point $\boldsymbol{w}_o$. Substituting $\boldsymbol{w}_o$ into (\ref{g3}), we obtain
\begin{equation}\label{gl}
	\boldsymbol{g}^{o}_t(i)=-cP_tE((\frac{c|\frac{e_o(i)}{\left \| \bar {\boldsymbol w}_o(i)\right \|}|^b}{|a-b|}+1)^\frac{a-b}{b}{|\frac{e_o(i)}{\left \| \bar {\boldsymbol w}_o(i)\right \|}|^{b-2}}\psi_o (i))
\end{equation}
where
\begin{equation}
	\psi_o (i)=\frac{\tilde {\boldsymbol{x}}(i)e_o(i)+\frac{e_o^2(i)\boldsymbol{w}_o(i)}{\|\bar {\boldsymbol w}_o(i)\|^2}}{\|\bar {\boldsymbol w}_o(i)\|^2}
\end{equation}
and $e_o(i)=v(i)-\boldsymbol{w}_o^T\boldsymbol{u}(i)$, $\|\tilde{\boldsymbol{w}}_o\|^2=\|\boldsymbol{w}_o\|^2+\phi$.

\textbf{Theorem 1}: Since $A=(\frac{c|\frac{e_o(i)}{\left \| \bar {\boldsymbol w}_o(i)\right \|}|^b}{|a-b|}+1)^\frac{a-b}{b}{|\frac{e_o(i)}{\left \| \bar {\boldsymbol w}_o(i)\right \|}|^{b-2}}$ and $B=\psi_o (i)$ are uncorrelated, (\ref{gl}) can be rewritten as
\begin{equation}
	\begin{aligned}
		\boldsymbol{g}^{o}_t(i)&=-cP_tE[((\frac{c|\frac{e_o(i)}{\left \| \bar {\boldsymbol w}_o(i)\right \|}|^b}{|a-b|}+1)^\frac{a-b}{b}{|\frac{e_o(i)}{\left \| \bar {\boldsymbol w}_o(i)\right \|}|^{b-2}})]E[\psi_o (i)]\\
		&=0
	\end{aligned}
\end{equation}

Thus, we can confirm that $\boldsymbol{w}_o$ is a local extremum point.

\textbf{Proof}:
The proof of the uncorrelation between $A$ and $B$ is equivalent to establishing $\operatorname{cov}(A, B) = 0$, where $\operatorname{cov}(\cdot,\cdot)$ denotes the covariance operation. An analysis of $B$ reveals that this is equivalent to investigating the correlation between $A$ and the components $\boldsymbol{x}(i)$, $v(i)$, and $\boldsymbol{u}(i)$.

Under A1, $A$ is uncorrelated with $\boldsymbol{x}(i)$. It remains to show that $A$ is also uncorrelated with both $v(i)$ and $\boldsymbol{u}(i)$, i.e.,
\begin{equation}
	\operatorname{cov}(A, v(i)) = \operatorname{cov}(A,\boldsymbol{u}(i)) = 0
\end{equation}

Given that $E(A)E(v(i)) = E(A)E(\boldsymbol{u}(i)) = 0$, the task simplifies to proving
\begin{equation}
	E(Av(i)) = E(A\boldsymbol{u}(i)) = 0
\end{equation}

Then, we have
\begin{equation}\label{21}
	\begin{aligned}
	&E(Av(i))=\frac{\alpha}{2\beta\gamma(1/\alpha)}\times\\
	&\int_{-\infty }^{+\infty}\exp[-(\frac{|\boldsymbol{u}(i)|}{\beta})^\alpha]\int_{-\infty }^{+\infty}A\exp[-(\frac{|v(i)|}{\beta})^\alpha]vdvdu
   \end{aligned}
\end{equation}

By utilizing the property that the integral of an odd function over a symmetric interval equals zero, Eq. (\ref{21}) is found to be 0. Similarly, it can be concluded that $E(A\boldsymbol{u}(i))=0$

Subsequently, we need to prove that $E(B)=0$, which can be expressed as
\begin{equation}
	\begin{aligned}
    E(B)&=\frac{\tilde {\boldsymbol{x}}(i)e_o(i)+\frac{e_o^2(i)\boldsymbol{w}_o(i)}{\|\bar {\boldsymbol w}_o(i)\|^2}}{\|\bar {\boldsymbol w}_o(i)\|^2}\\
    &=E(-\frac{\boldsymbol{u}(i)\boldsymbol{w}_o^T\boldsymbol{u}(i)}{\|\bar{\boldsymbol{w}}_o\|^2}+\frac{[v^2(i)+\boldsymbol{w}_o^T\boldsymbol{u}(i)\boldsymbol{w}_o^T\boldsymbol{u}(i)]\boldsymbol{w}_o}{\|\bar{\boldsymbol{w}}_o\|^4})\\
    &=-\frac{\sigma_i^2\boldsymbol{w}_o}{\|\bar{\boldsymbol{w}}_o\|^2}+\frac{\sigma_i^2\boldsymbol{w}_o}{\|\bar{\boldsymbol{w}}_o\|^2}\\
    &=0
    \end{aligned}
\end{equation}
This completes the proof.

\textbf{Theorem 2}:
Since the Hessian matrix $\boldsymbol{H}(\boldsymbol{w}_o)$ is positive definite, $\boldsymbol{w}_o$  is the local minimum point.

\textbf{Proof}:
The $\boldsymbol{H}(i)$ can be derived as
\begin{equation}
	\begin{aligned}\label{23}
				\boldsymbol{H}(i)&=P_t\frac{\partial \boldsymbol{g}(i)}{\partial \boldsymbol{w}^T(i)}\\
		&=P_tE[\hat{\boldsymbol{H}}_{1,A}(i)\hat{\boldsymbol{H}}_{1,B}(i))+\hat{\boldsymbol{H}}_{2,A}(i)\hat{\boldsymbol{H}}_{2,B}(i)]
	\end{aligned}
\end{equation}
where
\begin{equation}
	\begin{aligned}
		\hat{\boldsymbol{H}}_{1,A}(i)&=\left[c(\frac{c|\frac{e(i)}{\|\bar{\boldsymbol{w}}(i)\|}|^b}{|a-b|}+1)^\frac{a-2b}{b}\frac{c(a-b)}{|a-b|}|\frac{e(i)}{\|\bar{\boldsymbol{w}}(i)\|}|^{2b-4}\right.\\
		&\left.+c(\frac{c|\frac{e(i)}{\|\bar{\boldsymbol{w}}(i)\|}|^b}{|a-b|}+1)^\frac{a-b}{b}(b-2)|\frac{e(i)}{\|\bar{\boldsymbol{w}}(i)\|}|^{b-4}\right]
	\end{aligned}
\end{equation}
\begin{equation}
     \begin{aligned}
     \hat{\boldsymbol{H}}_{1,B}(i)&=\left[e^2(i)\tilde {\boldsymbol x}(i)\tilde {\boldsymbol x}(i)^T+\frac{\tilde {\boldsymbol x}(i)e^3(i)\boldsymbol w(i)^T}{\|\bar{\boldsymbol{w}}(i)\|^2}\right.\\
     	&\left. + \frac{e^3(i)\boldsymbol w(i)\tilde {\boldsymbol x}(i)^T}{\|\bar{\boldsymbol{w}}(i)\|^2} +\frac{e^4(i)\boldsymbol w(i)\boldsymbol w(i)^T}{\|\bar{\boldsymbol{w}}(i)\|^4}\right]/{\|\bar{\boldsymbol{w}}(i)\|^4}
     \end{aligned}
\end{equation}
\begin{equation}
	\hat{\boldsymbol{H}}_{2,A}(i)=c(\frac{c|\frac{e(i)}{\|\bar{\boldsymbol{w}}(i)\|}|^b}{|a-b|}+1)^\frac{a-b}{b}|\frac{e(i)}{\|\bar{\boldsymbol{w}}(i)\|}|^{b-2}
\end{equation}
\begin{equation}
	\begin{aligned}
		\hat{\boldsymbol{H}}_{2,B}(i)&=\left[\tilde {\boldsymbol x}(i)\tilde {\boldsymbol x}(i)^T\|\bar{\boldsymbol{w}}(i)\|^2+2e(i)\boldsymbol{w}(i)\tilde {\boldsymbol x}(i)^T-e^2(i)\textbf{I}\right.\\
		&\left.+\frac{4e^2(i)\boldsymbol{w}(i)\boldsymbol{w}(i)^T}{\|\bar{\boldsymbol{w}}(i)\|^2}+2e(i)\tilde {\boldsymbol x}(i)\boldsymbol{w}(i)^T\right]/\|\bar{\boldsymbol{w}}(i)\|^4
	\end{aligned}
\end{equation}

From \text{Theorem 1}, at $\boldsymbol{w}(i)=\boldsymbol{w}_o$, $\hat{\boldsymbol{H}}_{1,A}^o$ and $\hat{\boldsymbol{H}}_{1,B}^o$ are uncorrelated, and similarly, $\hat{\boldsymbol{H}}_{2,A}^o$ and $\hat{\boldsymbol{H}}_{2,B}^o$ are uncorrelated. Consequently, (\ref{23}) can be rewritten as 
\begin{equation}
		\boldsymbol{H}(\boldsymbol{w}_o)
		=P_tE[\hat{\boldsymbol{H}}_{1,A}^o]E[\hat{\boldsymbol{H}}_{1,B}^o]+P_tE[\hat{\boldsymbol{H}}_{2,A}^o]E[\hat{\boldsymbol{H}}_{2,B}^o]
\end{equation}
where
\begin{equation}
	\begin{aligned}
		E(\hat{\boldsymbol{H}}_{1,B}^o)&\approx\left[(\textbf{R}\|\bar{\boldsymbol{w}}_o\|^2\sigma_i^2+\sigma_i^4\|\bar{\boldsymbol{w}}_o\|^2+2\sigma_i^4{\boldsymbol{w}}_o{\boldsymbol{w}}_o^T)\right.\\
		&\left.-3\sigma_i^4{\boldsymbol{w}}_o{\boldsymbol{w}}_o^T-3\sigma_i^4{\boldsymbol{w}}_o{\boldsymbol{w}}_o^T+3\sigma_i^4{\boldsymbol{w}}_o{\boldsymbol{w}}_o^T\right]/\|\bar{\boldsymbol{w}}_o\|^4\\
		&=\frac{\textbf{R}\sigma_i^2}{\|\bar{\boldsymbol{w}}_o\|^2}+\frac{\sigma_i^4\text{I}}{\|\bar{\boldsymbol{w}}_o\|^2}-\frac{\sigma_i^4\boldsymbol{w}_o{\boldsymbol{w}}_o^T}{\|\bar{\boldsymbol{w}}_o\|^4}
	\end{aligned}
\end{equation}
\begin{equation}
	\begin{aligned}
		E(\hat{\boldsymbol{H}}_{2,B}^o)&=\left[(\textbf{R}\|\bar{\boldsymbol{w}}_o\|^2+\sigma_i^2\|\bar{\boldsymbol{w}}_o\|^2\textbf{I})-2\sigma_i^2\boldsymbol{w}_o{\boldsymbol{w}}_o^T\right.\\
		&\left.-\sigma_i^2\|\bar{\boldsymbol{w}}_o\|^2\textbf{I}+4\sigma_i^2\boldsymbol{w}_o{\boldsymbol{w}}_o^T-2\sigma_i^2\boldsymbol{w}_o{\boldsymbol{w}}_o^T\right]/\|\bar{\boldsymbol{w}}_o\|^4\\
		&=\frac{\textbf{R}}{\|\bar{\boldsymbol{w}}_o\|^2}
	\end{aligned}
\end{equation}

It can be readily observed that $\hat{\boldsymbol{H}}_{1,B}^o$, $\hat{\boldsymbol{H}}_{2,A}^o$ and $\hat{\boldsymbol{H}}_{2,B}^o$ are all strictly positive, whereas the sign of $\hat{\boldsymbol{H}}_{1,A}^o$ depends on parameters. Assuming the variance of the input noise is sufficiently small, we have $\hat{\boldsymbol{H}}_{2,B}^o \gg  \hat{\boldsymbol{H}}_{2,A}^o$ which guarantees the positive definiteness of the Hessian matrix $\boldsymbol{H}(\boldsymbol{w}_o)$. Thus, $\boldsymbol{w}_o$  is the local minimum point.

This completes the proof.

\textbf{Remark} \textbf{4}: It is noteworthy that Gaussian and generalized Gaussian variables with equal variance possess identical second-order moments but differ in their fourth-order moments.   As noted above, the input noise variance is generally small; thus, we neglect the residual terms arising from differences in higher-order moments, leading to the simplified expression denoted as $\hat{\boldsymbol{H}}_{1,B}^o$.

\subsection{Mean Convergence}
This subsection presents a stability analysis of the RTGA-OC algorithm, deriving the range of step sizes that ensures mean convergence.

\textbf{Theorem 3}: The mean stability of the RTGA-OC algorithm can be guaranteed by
\begin{equation}
	\begin{aligned}
&0<\mu<\\
&2/\{P_t((\frac{c^2(a-b)}{|a-b|}\vartheta_o ^{2b-4}+c(b-2)\vartheta_o^{b-4})\lambda_{\max}[E(\hat{\boldsymbol{H}}_{1,B}^o)]\\
&+c\vartheta_o^{b-2}\lambda_{\max}[E(\hat{\boldsymbol{H}}_{2,B}^o)])\}
	\end{aligned}
\end{equation}

\textbf{Proof}: 
The gradient error occurs because the expected value in the weight update formula (\ref{update}) is replaced by its instantaneous counterpart. The gradient error is defined as
\begin{equation}\label{38}
	\boldsymbol \varepsilon (i) = \hat{\boldsymbol{g}}(i) - {\boldsymbol{g}}(i)
\end{equation}

Substituting (\ref{38}) into (\ref{update}), gives
\begin{equation}\label{39}
	\boldsymbol{w}(i+1)=\boldsymbol{w}(i)-\mu {\boldsymbol{g}}(i)-\mu \boldsymbol \varepsilon (i)
\end{equation}

By defining $\Delta  \boldsymbol{w}(i)= \boldsymbol {w}_o -\boldsymbol {w}(i)$, and introducing it into (\ref{39}), yields
\begin{equation}\label{40}
	\Delta  \boldsymbol{w}(i+1)=\Delta  \boldsymbol{w}(i)+\mu[\boldsymbol \varepsilon (i)+\boldsymbol{g}(i)]
\end{equation}

Using $\boldsymbol{g}(i)\approx -\mu\boldsymbol{H}(\boldsymbol{w}_o)\Delta  \boldsymbol{w}(i)$ \cite{kelley1999iterative}, (\ref{40}) can be further obtained as
\begin{equation}\label{41}
	\Delta  \boldsymbol{w}(i+1)=(\textbf{I}-\mu\boldsymbol{H}(\boldsymbol{w}_o))\Delta  \boldsymbol{w}(i)+\mu\boldsymbol \varepsilon (i)
\end{equation}

Taking the expectation of both sides of (\ref{41}), gives
\begin{equation}
	E(	\Delta  \boldsymbol{w}(i+1))=(\textbf{I}-\mu\boldsymbol{H}(\boldsymbol{w}_o))E(\Delta  \boldsymbol{w}(i))
\end{equation}

From (33), ensuring the local convergence of the RTGA-OC algorithm requires that all eigenvalues of matrix $\mathbf{H}(\boldsymbol{w}_o))$ satisfy
\begin{equation}
	|\textbf{I} - \mu \lambda_{\max}[\mathbf{H}(\boldsymbol{w}_o)]|<1
\end{equation}
which can be simplified as
\begin{equation}
	0<\mu<\frac{2}{\lambda_{\max}[\mathbf{H}(\boldsymbol{w}_o)]}
\end{equation}
where $\lambda_{\max}[\cdot]$ is maximum eigenvalue of the matrix.

The determination of $\mathbf{H}(\boldsymbol{w}_o)$ requires the prior evaluation of $E[\hat{\boldsymbol{H}}_{1,A}^o]$ and $E[\hat{\boldsymbol{H}}_{2,A}^o]$.
When the algorithm reaches steady state, $e_o(i)$ becomes sufficiently small, the nonlinear terms in $\hat{\boldsymbol{H}}_{1,A}^o$ and $\hat{\boldsymbol{H}}_{2,A}^o$ are simplified using the Taylor approximation as
\begin{equation}
	c(\frac{c|\frac{e_o}{\|\bar{\boldsymbol{w}}_o\|}|^b}{|a-b|}+1)^\frac{a-2b}{b}\approx 1+\frac{a-2b}{b}\frac{c|\frac{e_o}{\|\bar{\boldsymbol{w}}_o\|}|^b}{|a-b|}
\end{equation}
\begin{equation}
	c(\frac{c|\frac{e_o}{\|\bar{\boldsymbol{w}}_o\|}|^b}{|a-b|}+1)^\frac{a-b}{b}\approx 1+\frac{a-b}{b}\frac{c|\frac{e_o}{\|\bar{\boldsymbol{w}}_o\|}|^b}{|a-b|}
\end{equation}

Then, we have
\begin{equation}
	\begin{aligned}\label{41H}
		E[\hat{\boldsymbol{H}}_{1,A}^o]&=\frac{c^2(a-b)}{|a-b|}\vartheta_o ^{2b-4}+\frac{c^3(a-2b)}{b(a-b)}\vartheta_o^{3b-4}\\
		&+c(b-2)\vartheta_o^{b-4}+\frac{c^2(b-2(a-b))}{b|a-b|}\vartheta_o^{2b-4}
	\end{aligned}
\end{equation}
\begin{equation}\label{42H}
	E[\hat{\boldsymbol{H}}_{2,A}^o]=c\vartheta_o^{b-2}+\frac{c^2(a-b)}{b|a-b|}\vartheta_o^{2b-2}
\end{equation}
where
\begin{equation}
	\vartheta_o=E|\frac{e_o}{\|\bar{\boldsymbol{w}}_o\|}|
\end{equation}
\begin{equation}
	\vartheta_o^m=\frac{(\sigma_i\|\bar{\boldsymbol{w}}_o\|)^m\frac{\Gamma (\frac{m+1}{\alpha})}{\Gamma (\frac{1}{\alpha })}(\frac{\Gamma (\frac{1}{\alpha})}{\Gamma (\frac{3}{\alpha })})^{\frac{m}{2}}}{\|\bar{\boldsymbol{w}}_o\|^m}
\end{equation}
and $\Gamma(\cdot)$ is the gamma function.

\textbf{Remark} \textbf{5}: By examining Eqs. (\ref{41H}) and (\ref{42H}), it can be observed that when approximated to the second order, certain parameter choices may cause $E[\hat{\boldsymbol{H}}_{2,A}^o]<0
$, thereby compromising the stability of the Hessian matrix. To avoid this issue, we choose to expand the nonlinear term only to the first Taylor order, which leads to the following revised expressions for the $	E[\hat{\boldsymbol{H}}_{1,A}^o]$ and $	E[\hat{\boldsymbol{H}}_{1,A}^o]$
as 
\begin{equation}
		E[\hat{\boldsymbol{H}}_{1,A}^o]=\frac{c^2(a-b)}{|a-b|}\vartheta_o ^{2b-4}+c(b-2)\vartheta_o^{b-4}
\end{equation}
\begin{equation}
	E[\hat{\boldsymbol{H}}_{2,A}^o]=c\vartheta_o^{b-2}
\end{equation}

This completes the proof.
\subsection{Steady-State MSD}
This subsection presents the steady-state mean-square deviation (MSD) analysis of the RTGA-OC algorithm under generalized Gaussian noise conditions.

\textbf{Theorem 4}: The steady-state MSD can be derived as
\begin{equation}
				E[\|\Delta  \boldsymbol{w}(\infty)\|^2]\approx \mu^2\boldsymbol{s}^T(\textbf{I}-\boldsymbol{F})^{-1}\text{vec}\{\textbf{I}\}
\end{equation}

\textbf{Proof}: As $\boldsymbol g(i)$ approaches zero in steady state, from (\ref{38}), we can get   $\boldsymbol \varepsilon(i) \approx \hat {\boldsymbol{g}}(i)$. By squaring both sides of Eq. (\ref{41}) and taking the Euclidean norm, we obtain
\begin{equation}\label{47}
	E[\|\Delta  \boldsymbol{w}(i+1)\|^2_{\boldsymbol Z}]\approx E[\|\Delta  \boldsymbol{w}(i)\|^2_{\boldsymbol N}]+\mu^2E[\|\boldsymbol \varepsilon(i)\|^2_{\boldsymbol Z}]
\end{equation}
where $\boldsymbol{Z}$ is a freely selectable positive semi-definite matrix, and $\boldsymbol{N}$ is defined as
\begin{equation}\label{477}
   \boldsymbol{N}=(\textbf{I}-\mu\boldsymbol{H}(\boldsymbol{w}_o))\boldsymbol{Z}(\textbf{I}-\mu\boldsymbol{H}(\boldsymbol{w}_o))
\end{equation}

For the purpose of the subsequent analysis, we define $\boldsymbol z=\text{vec}(\boldsymbol Z)$; $\boldsymbol s=\text{vec}(\boldsymbol S)$; $\boldsymbol n=\text{vec}(\boldsymbol N)$, where $\text{vec}(\cdot)$ denotes the vectorization operator and $\boldsymbol{S}$ is defined as
\begin{equation}
	\begin{aligned}
		\boldsymbol{S}&=E[\boldsymbol \varepsilon(\boldsymbol{w}_o)\boldsymbol \varepsilon(\boldsymbol{w}_o)^T]\\
		&=c^2\vartheta ^{2b-4}_o[\frac{\textbf{R}\sigma_i^2}{\|\bar{\boldsymbol{w}}_o\|^2}
		+\frac{\sigma_i^4\text{I}}{\|\bar{\boldsymbol{w}}_o\|^2}-\frac{\sigma_i^4\boldsymbol{w}_o{\boldsymbol{w}}_o^T}{\|\bar{\boldsymbol{w}}_o\|^4}]
	\end{aligned}
\end{equation}

Subsequently, using the relationship between the vectorization operator and the matrix trace \cite{abadir2005matrix}, and noting that $\boldsymbol{Z}$ is symmetric and deterministic, we obtain
\begin{equation}\label{49}
	\begin{aligned}
	E[ \| \boldsymbol{\varepsilon}(\boldsymbol{w}_o)\|^2_{\boldsymbol Z}]&=E[\text{tr}\{{\boldsymbol Z}(\boldsymbol \varepsilon(\boldsymbol{w}_o)\boldsymbol \varepsilon(\boldsymbol{w}_o)^T)\}]\\
	&=\text{tr}\{{\boldsymbol Z}{\boldsymbol S}\}\\
	&={\boldsymbol s}^T{\boldsymbol z}
    \end{aligned}
\end{equation}
where $\text{tr}\{\cdot\}$ calculates the trace of matrix.

Substituting (\ref{49}) into (\ref{47}), gives
\begin{equation}\label{50}
		E[\|\Delta  \boldsymbol{w}(i+1)\|^2_{\boldsymbol z}]\approx E[\|\Delta  \boldsymbol{w}(i)\|^2_{\boldsymbol n}]+\mu^2\boldsymbol{s}^T\boldsymbol{z}
\end{equation}

For any matrices $\mathcal{A}$, $\mathcal{B}$, and $\mathcal{C}$, the following identity holds: $\operatorname{vec}(\mathcal{A}\mathcal{B}\mathcal{C}) = (\mathcal{C}^T \otimes \mathcal{A}) \operatorname{vec}(\mathcal{B})$. Applying this property, (\ref{477}) can be rewritten as
\begin{equation}\label{51}
	\boldsymbol{n}=\boldsymbol{F}\boldsymbol{z}
\end{equation}
where 
\begin{equation}
	\boldsymbol{F}\triangleq (\textbf{I}-\mu\boldsymbol{H}(\boldsymbol{w}_o))\otimes(\textbf{I}-\mu\boldsymbol{H}(\boldsymbol{w}_o))
\end{equation}
and $\otimes$ represents the Kronecker product.

In steady state, substituting (\ref{51}) into (\ref{50}), obtains
\begin{equation}
		E[\|\Delta  \boldsymbol{w}(\infty)\|^2_{(\textbf{I}-\boldsymbol F )\boldsymbol z}]\approx \mu^2\boldsymbol{s}^T\boldsymbol{z}
\end{equation}

After selecting $\boldsymbol{z}$ as $(\textbf{I}-\boldsymbol{F})^{-1}\text{vec}\{\textbf{I}\}$, the steady-state MSD can be express as
\begin{equation}
			E[\|\Delta  \boldsymbol{w}(\infty)\|^2]\approx \mu^2\boldsymbol{s}^T(\textbf{I}-\boldsymbol{F})^{-1}\text{vec}\{\textbf{I}\}
\end{equation}

This completes the proof.

\section{Complexity of the Calculation}
\begin{table*}[!t]
	\centering
	\small
	\caption{Computation Complexity}
	\label{tab:complexity}
	\begin{tabular}{lccc}
		\toprule
		Algorithm & + / -  & $\times$ / $\div$ & Nonlinear \\
		\midrule
		GMCC \cite{GMCC}      & $2L$   & $2L+2+2\alpha$               & 4         \\
		RGA  \cite{RGA}      & $2L+4$ & $2L+1+2\beta+|\alpha|/\beta$  &4          \\
		GDTLS \cite{GDTLS}     & $4L$   & $5L+5$                        & 0         \\
		MTC \cite{MTC}       & $4L$   & $5L+11$                       & 1         \\
		MTGC \cite{MTGC}     & $4L$   & $5L+7+2\alpha$                & 4         \\
		GMBZTC \cite{GMBZTC}    & $4L+1$ & $5L+7+2\alpha$                & 5         \\
		TACLMD \cite{TACLDM}    & $4L+2$ & $5L+18$                       & 3         \\ 
    	RTGA   & $4L+4$ & $5L+5+2b+|a|/b$               &3           \\ 
		Proposed   & $(4L+4)(1-P_{ce})(L_\text{reused}+P_{ce}L_\text{reused}+1)$ & $(5L+5+2b+|a|/b)(1-P_{ce})(L_\text{reused}+P_{ce}L_\text{reused}+1)$               &3           \\
		\bottomrule
	\end{tabular}
\end{table*}
This subsection presents a computational complexity analysis of the considered algorithms, with the results summarized in Table I. 

As illustrated in Table IV, the proposed RTGA algorithm manifests a slightly higher computational complexity than the MTGC, GMBZTC, and TACLDM algorithms when implemented without IDR and OC methods. It is worth noting, however, that the complexity increase depends on parameters, specifically the values of $a$ and $b$. For small values of $a$ and $b$, the complexity of RTGA remains nearly comparable to that of the other algorithms. Furthermore, when $\alpha = 2$, the MTGC algorithm exhibits exactly the same complexity as the MTC algorithm. When the IDR and OC strategies are employed, the resulting complexity is influenced by the values of $P_{ce}$ and $L_{\text{reused}}$. In applications demanding fast convergence, such as AEC, the joint use of IDR and OC not only accelerates convergence but also restrains computational load by selectively limiting updates. In particular, when $L_{\text{reused}} = 1$ and $P_{ce} = 70\%$, the proposed IDROC variant is expected to achieve computational complexity comparable to that of the original algorithm, with this point being verified in the simulations section.
\section{Simulation}
This section presents the simulation results for system identification and AEC. All experiments, with the exception of AEC, are performed using 1000 independent Monte Carlo runs. Unless specified otherwise, the unknown weight vectors are randomly generated, with a filter order $L=9$ and 8000 sampling points. Furthermore, to guarantee a fair comparison, the step sizes of the different algorithms are set such that all algorithms exhibit identical initial convergence rates. The performance of all algorithms is evaluated using the normalized MSD (NMSD), which is defined as
\begin{equation}
	\text{NMSD(dB)}=10\log_{10}E[\|\boldsymbol{w}(i)-\boldsymbol{w}_o\|^2/\|\boldsymbol{w}_o\|^2]
\end{equation}

\subsection{System identification}
In this subsection, the noise cases used in the subsequent simulations are described as follows:
\begin{itemize}
	\item Case 1 (Gaussian noise): Both the input noise $\boldsymbol{u}(i)$ and the output noise $v(i)$ are zero-mean Gaussian processes with a variance of 0.1.
	\item Case 2 (Impulsive noise): The input noise $\boldsymbol{u}(i)$ is a zero-mean Gaussian process with a variance of 0.1. The output noise $v(i)$ is a mixture of a zero-mean Gaussian process (variance of 0.1) and an impulsive noise process (variance of 100), generated by a Bernoulli process \cite{MTC}.
	\item Case 3 (Laplace noise): The input noise $\boldsymbol{u}(i)$ is a zero-mean Gaussian process with a variance of 0.1, while the output noise $v(i)$ is zero-mean Laplace noise with a variance of 1.
	\item Case 4 (Uniform noise): The input noise $\boldsymbol{u}(i)$ is zero-mean uniform noise with a variance of 1. The output noise $v(i)$ is a mixture of zero-mean uniform noise (variance of 1) and impulsive noise (variance of 100).
	\item Case 5 (Binary noise): The input noise $\boldsymbol{u}(i)$ is zero-mean binary noise with a variance of 0.2. The output noise $v(i)$ is a mixture of zero-mean binary noise (variance of 0.2) and impulsive noise (variance of 100).
\end{itemize}
Furthermore, Table II summarizes the parameter settings employed by the different algorithms across the various noise cases.
\begin{table}[!t]
	\centering
	\scriptsize
	\setlength{\tabcolsep}{2.5pt}
	\caption{Parameter Settings for Different Cases}
	\label{tab:parameters}
	\begin{tabular}{lccccc}
		\toprule
		Algorithm & Case 1  & Case 2  & Case 3 & Case 4 & Case 5\\
		\midrule
		GMCC & 
		\makecell[c]{$\mu=0.004$ \\ $\alpha=2$ \\ $\sigma=1$} & 
		\makecell[c]{$\mu=0.006$ \\ $\alpha=2$ \\ $\sigma=1$} & 
		\makecell[c]{$\mu=0.1$ \\ $\alpha=1$ \\ $\sigma=1$} & 
		\makecell[c]{$\mu=0.033$ \\ $\alpha=6$ \\ $\sigma=1$} & 
		\makecell[c]{$\mu=0.1$ \\ $\alpha=4$ \\ $\sigma=1$} \\
		\cmidrule(r){1-6}
		
		RGA & 
		\makecell[c]{$\mu=0.4$ \\ $\alpha=-100$ \\ $\beta=2.1$ \\ $\lambda=0.01$} & 
		\makecell[c]{$\mu=0.4$ \\ $\alpha=-100$ \\ $\beta=2.1$ \\ $\lambda=0.01$} & 
		\makecell[c]{$\mu=2.75$ \\ $\alpha=-100$ \\ $\beta=1.5$ \\ $\lambda=0.01$} & 
		\makecell[c]{$\mu=0.25$ \\ $\alpha=-1000$ \\ $\beta=8$ \\ $\lambda=0.01$} & 
		\makecell[c]{$\mu=0.03$ \\ $\alpha=-100$ \\ $\beta=2.3$ \\ $\lambda=1$} \\
		\cmidrule(r){1-6}
		
		GDTLS & $\mu=0.0022$ & $\mu=0.0022$ & $\mu=0.25$ & $\mu=0.12$ & $\mu=0.05$ \\
		\cmidrule(r){1-6}
		
		MTC 
		& \makecell[c]{$\mu=0.003$ \\ $\sigma=1$}
		 & \makecell[c]{$\mu=0.003$  \\ $\sigma=1$}
		  & \makecell[c]{$\mu=0.26$  \\ $\sigma=1$}
		   & \makecell[c]{$\mu=0.17$  \\ $\sigma=1$}
		    & \makecell[c]{$\mu=0.03$  \\ $\sigma=1$} \\
		\cmidrule(r){1-6}
		
		MTGC & 
		\makecell[c]{$\mu=0.003$ \\ $\alpha=2$ \\ $\sigma=1$} & 
		\makecell[c]{$\mu=0.006$ \\ $\alpha=2$ \\ $\sigma=1$} & 
		\makecell[c]{$\mu=0.18$ \\ $\alpha=1.56$ \\ $\sigma=1$} & 
		\makecell[c]{$\mu=0.055$ \\ $\alpha=6$ \\ $\sigma=1$} & 
		\makecell[c]{$\mu=0.05$ \\ $\alpha=2.34$ \\ $\sigma=1$} \\
		\cmidrule(r){1-6}
		
		GMBZTC & 
		\makecell[c]{$\mu=0.005$ \\ $\alpha=2$ \\ $\sigma=1$ \\ $\gamma=1$} & 
		\makecell[c]{$\mu=0.01$ \\ $\alpha=2$ \\ $\sigma=1$ \\ $\gamma=1$} & 
		\makecell[c]{$\mu=0.14$ \\ $\alpha=1.56$ \\ $\sigma=1$ \\ $\gamma=0.1$} & 
		\makecell[c]{$\mu=0.046$ \\ $\alpha=6$ \\ $\sigma=1$ \\ $\gamma=0.51$} & 
		\makecell[c]{$\mu=0.07$ \\ $\alpha=2.34$ \\ $\sigma=1$ \\ $\gamma=0.91$} \\
		\cmidrule(r){1-6}
		
		TACLMD &	
			\makecell[c]{$\mu=0.005$ \\ $\gamma=1$} & 
		\makecell[c]{$\mu=0.105$  \\ $\gamma=1.4$} & 
		\makecell[c]{$\mu=2.4$ \\ $\gamma=1.3$} & 
		\makecell[c]{$\mu=5$ \\ $\gamma=4$} & 
		\makecell[c]{$\mu=0.085$ \\ $\gamma=0.72$} \\
		\cmidrule(r){1-6}
		
		RTGA & 
		\makecell[c]{$\mu=0.022$ \\ $a=-100$ \\ $b=2$ \\ $c=0.2$} & 
		\makecell[c]{$\mu=0.022$ \\ $a=-100$ \\ $b=2$ \\ $c=0.1$} & 
		\makecell[c]{$\mu=0.47$ \\ $a=-100$ \\ $b=1.5$ \\ $c=0.2$} & 
		\makecell[c]{$\mu=0.055$ \\ $a=-1000$ \\ $b=8$ \\ $c=0.6$} & 
		\makecell[c]{$\mu=0.02$ \\ $a=-100$ \\ $b=2.3$ \\ $c=1.5$} \\ 
		\cmidrule(r){1-6}
		
		\makecell[c]{Proposed\\$L_\text{reused}=3$} & 
		\makecell[c]{$\mu=0.0098$ \\ $a=-100$ \\ $b=2$ \\ $c=0.1$} & 
		\makecell[c]{$\mu=0.0088$ \\ $a=-100$ \\ $b=2$ \\ $c=0.18$} & 
		\makecell[c]{$\mu=0.155$ \\ $a=-100$ \\ $b=1.5$ \\ $c=0.1$} & 
		\makecell[c]{$\mu=0.025$ \\ $a=-1000$ \\ $b=8$ \\ $c=0.6$} & 
		\makecell[c]{$\mu=0.01$ \\ $a=-100$ \\ $b=2.29$ \\ $c=1.2$} \\
		\bottomrule
	\end{tabular}
\end{table}

\subsubsection{Proposed Threshold and Runtime}
\mbox{}

The experimental analysis under Case 1 employs the parameter settings listed in Table III. The validation of (\ref{6}) is conducted by computing the estimated censoring ratio \(\hat{P}_{ce}\), while the computational complexity reported in Table I is verified by comparing the algorithms' runtime. All quantitative results are consolidated in Table II. Importantly, the conventional method is used here instead of (\ref{8}) because it achieves better estimation accuracy under Gaussian noise conditions. As shown in Table II, the estimated censoring ratio \(\hat{P}_{ce}\) is in close agreement with the preset value, thereby validating the threshold calculation in (\ref{6}). Furthermore, when \(L_{\text{reused}} = 1\) and \(P_{ce} = 70\%\), the runtime of the proposed algorithm is comparable to that of the RTGA. The slight increase is attributable to the additional computation required for estimating \(\sigma_e\), which corroborates the computational complexity analysis of the proposed algorithm presented in Table I. 
\begin{table}[htbp]
	\caption{Quantitative results of different algorithm}
	\centering
	\small
	\begin{tabular}{lcccc}
		\toprule
		Algorithm & $L_\text{reused}$& $\hat{P}_{ce}$(\%)  &  Runtime(s)  & NMSD(dB) \\
		\midrule 
		GMCC      & -  & -       & 0.0041    & -19.6202         \\
		RGA      & -  & -       & 0.0065    & -19.4157         \\
		GDTLS     & -   & -      & 0.0102            & -29.5989        \\
		MTC       &-    & -     & 0.0101          & -28.8217         \\
		MTGC      & -  & -       &0.0102    & -28.8217         \\
		GMBZTC    & -  & -     & 0.0110   & -29.3629         \\
		TACLMD    & -   & -    &0.0106           & -29.3714     \\   
        RTGA    & -  & -     & 0.0104          & -29.6634     \\
       
        Proposed(30\%)  & 1  & 30.12      & 0.0180          & -31.2739     \\ 
       Proposed(50\%)   & 1 & 50.21    & 0.0145           & -31.2366     \\ 
       Proposed(70\%)   & 1 & 70.30      & 0.0113           & -31.2360     \\ 
        Proposed(30\%)   & 3 & 30.14     & 0.0304           & -29.9519     \\ 
        Proposed(50\%)  & 3  & 50.20      &0.0241           & -29.9607     \\ 
        Proposed(70\%)  & 3  & 70.28      & 0.0182           & -29.8918     \\ 
        
		\bottomrule
	\end{tabular}
\end{table}
\subsubsection{Impact of Parameter c and Reuse Times on Algorithm Performance}
\mbox{}

\begin{figure}[htbp]
	\centering
	\begin{minipage}{0.24\textwidth}
		\centering
		\includegraphics[width=\linewidth]{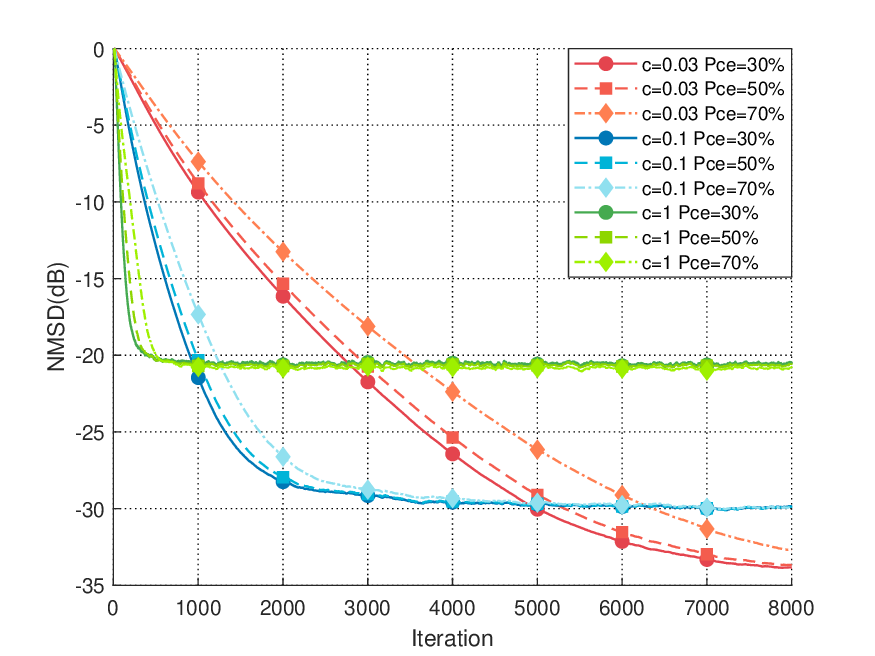}
		\\\text{(a)}
	\end{minipage}
	\hfill
	\begin{minipage}{0.24\textwidth}
		\centering
		\includegraphics[width=\linewidth]{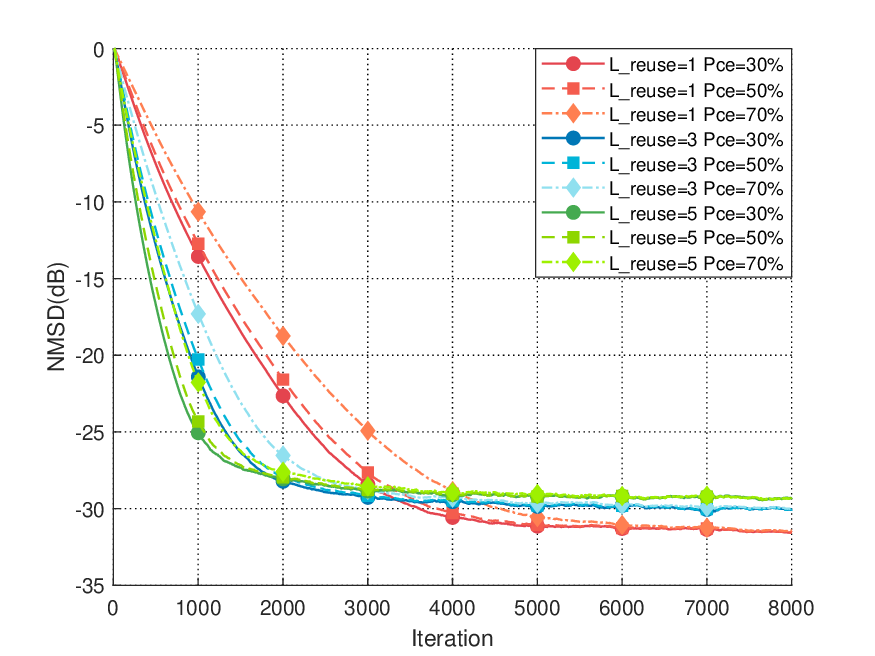}
		\\\text{(b)}
	\end{minipage}

	\caption{Performance curves of the RTGA-IDROC algorithm under varying parameters: (a) parameter c, and (b) reuse times $L_\text{reused}$}
	\label{fig4}
\end{figure}

The effects of parameters $a$ and $b$ on the proposed algorithm were discussed in detail previously. The impact of parameter $c$ and the reuse count on the RTGA-IDROC algorithm is now examined. The NMSD performance under different values of $c$ and various reuse counts is presented in Figs. \ref{fig4}(a) and (b), respectively. As indicated in Fig. \ref{fig4}(a), faster convergence speed is achieved at the cost of an increased steady-state error when $c$ is increased, whereas the opposite effect is observed when $c$ is reduced. It can be seen from Fig. \ref{fig4}(b) that the convergence speed is enhanced by increasing the reuse count, but this is accompanied by a rise in the steady-state error. Conversely, a reduction in the reuse count is found to produce the opposite outcome. Notably, since the improvement achieved by increasing the reuse count from 3 to 5 is considered marginal, a reuse count of 3 is adopted for the subsequent simulations.

\begin{figure}[htbp]
	\centering
	\begin{minipage}{0.24\textwidth}
		\centering
		\includegraphics[width=\linewidth]{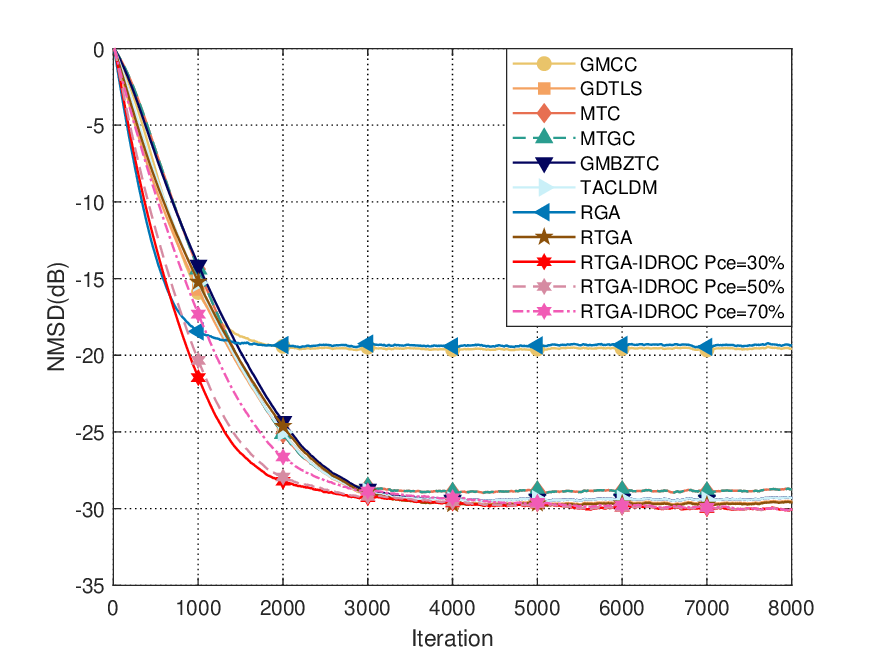}
		\\\text{(a)}
	\end{minipage}
	\hfill
	\begin{minipage}{0.24\textwidth}
		\centering
		\includegraphics[width=\linewidth]{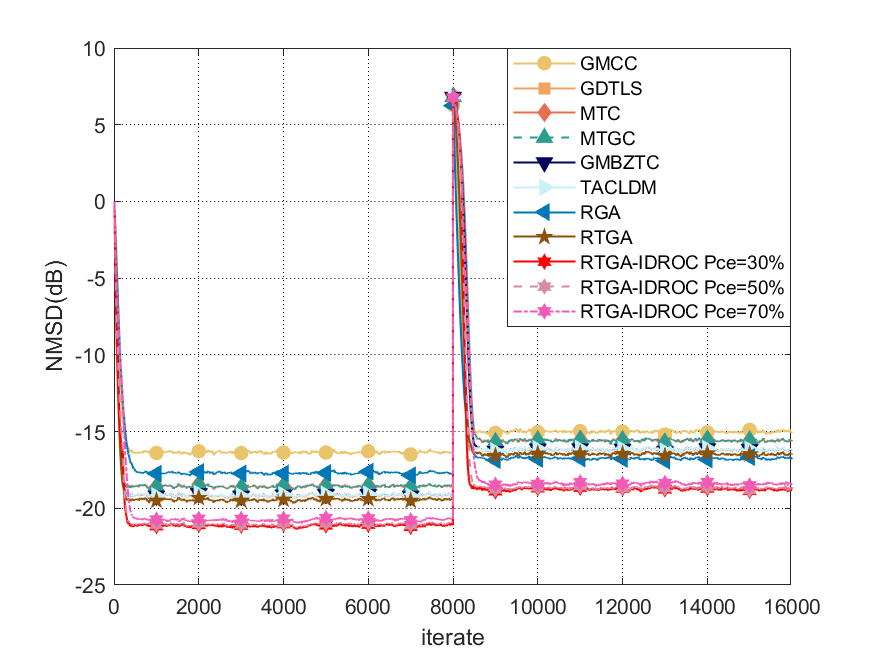}
		\\\text{(b)}
	\end{minipage}
	
	\caption{Performance comparison under Gaussian noise}
	\label{fig5}
\end{figure}

\subsubsection{Comparative Analysis and Tracking Performance in Case 1}
\mbox{}

Figs. \ref{fig5}(a) and 5(b) show the NMSD performance of the proposed algorithm and other competing algorithms under Gaussian noise, and the tracking performance when the impulse response of the unknown system suddenly shifts right by 3 samples at time $i=8000$, respectively. As depicted in Fig. \ref{fig5}(a), under Gaussian noise, the proposed algorithm achieves faster convergence speed and lower steady-state error than other algorithms at different censoring rates. Specifically, using -25dB as the baseline, RTGA-IDROC(30\%) requires 1310 iterations, RTGA-IDROC(50\%) requires 1410 iterations, RTGA-IDROC(70\%) requires 1750 iterations, while other TLS-based algorithms require over 2000 iterations. This improvement is attributed to the proposed IDR method, which enhances performance by selecting data points that are farther apart, thereby obtaining reused data with lower correlation. Notably, the accelerated convergence of RTGA-IDROC compared to RTGA is achieved without sacrificing steady-state accuracy, which validates the discussion in Section II-B.   Furthermore, the associated increase in computational complexity from the IDR method is constrained by the OC strategy. Results in Fig. \ref{fig5}(b) demonstrate that the tracking performance of the proposed algorithm is significantly better than that of the competing algorithms, which also benefits from the combined effect of the IDR and OC methods. It is important to note that although the IDR method improves performance by uniformly reusing past data, it has two limitations: 1) When the system changes, past input-output relationships may not reflect the current system; 2) In practical applications considered later, storing all past data may consume excessive memory resources. Therefore, in the tracking performance test and subsequent AEC application tests, the data reuse is limited to the most recent 200 points prior to the current time instant.

\subsubsection{Comparative Analysis in Cases 2--5}
\mbox{}

A comparison of the NMSD performance between the proposed algorithm and competing methods under various noise conditions is presented in Fig. 6. As shown, the RTGA-IDROC method is demonstrated to achieve significantly superior performance compared to other competing algorithms across all tested noise environments. Furthermore, the RTGA-IDROC algorithm is considerably enhanced through the integrated application of the IDR and OC strategies relative to the original RTGA algorithm. These results confirm the effectiveness of the RTGA approach while successfully validating the capability of the proposed IDROC strategy to improve algorithmic performance.
\begin{figure}[t!]
	\centering
	\begin{minipage}{0.24\textwidth}
		\centering
		\includegraphics[width=\linewidth]{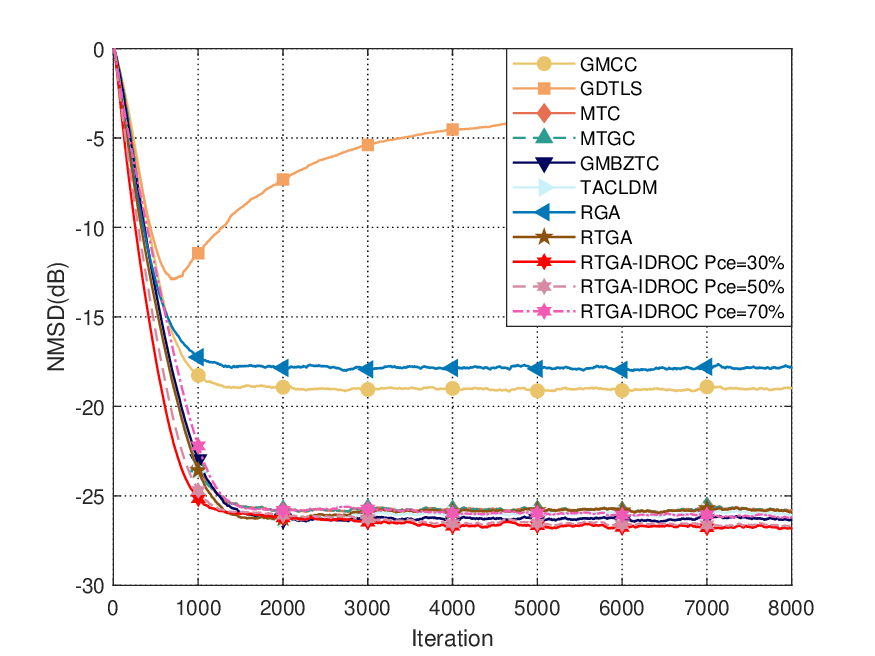}
		\\\text{(a)} 
	\end{minipage}
	\hfill
	\begin{minipage}{0.24\textwidth}
		\centering
		\includegraphics[width=\linewidth]{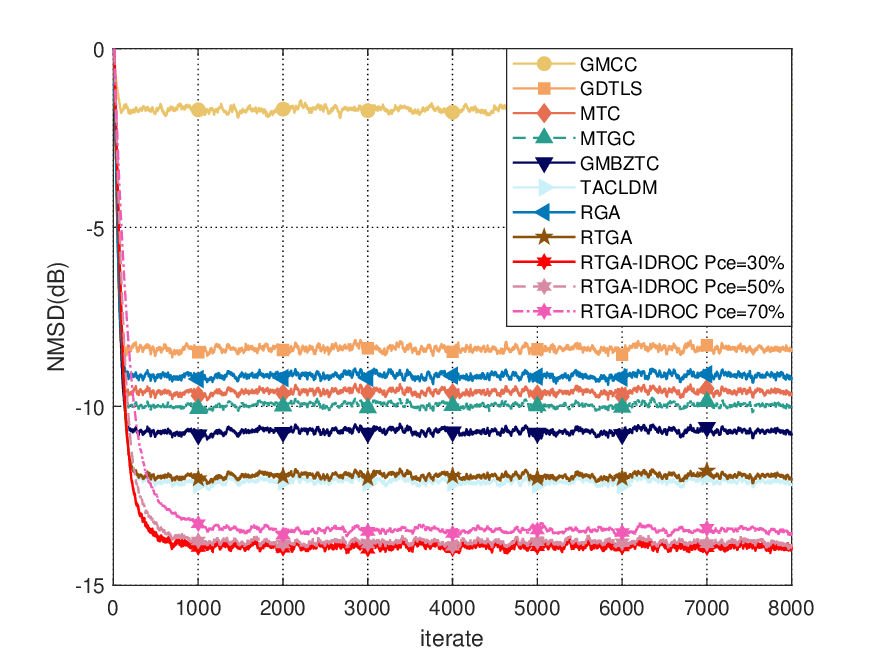}
		\\\text{(b)} 
	\end{minipage}
	
	\begin{minipage}{0.24\textwidth}
		\centering
		\includegraphics[width=\linewidth]{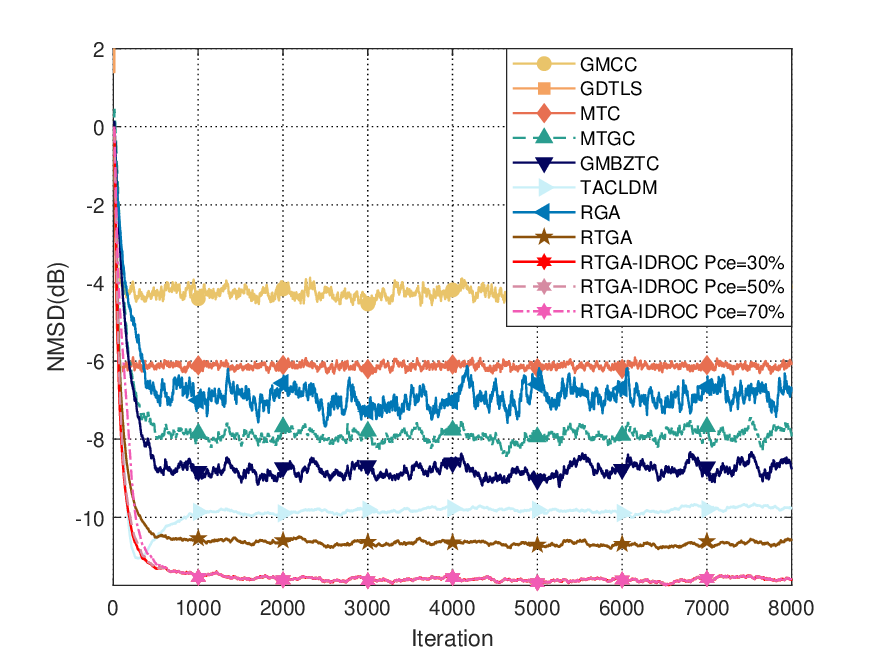}
		\\\text{(c)} 
	\end{minipage}
	\hfill
	\begin{minipage}{0.24\textwidth}
		\centering
		\includegraphics[width=\linewidth]{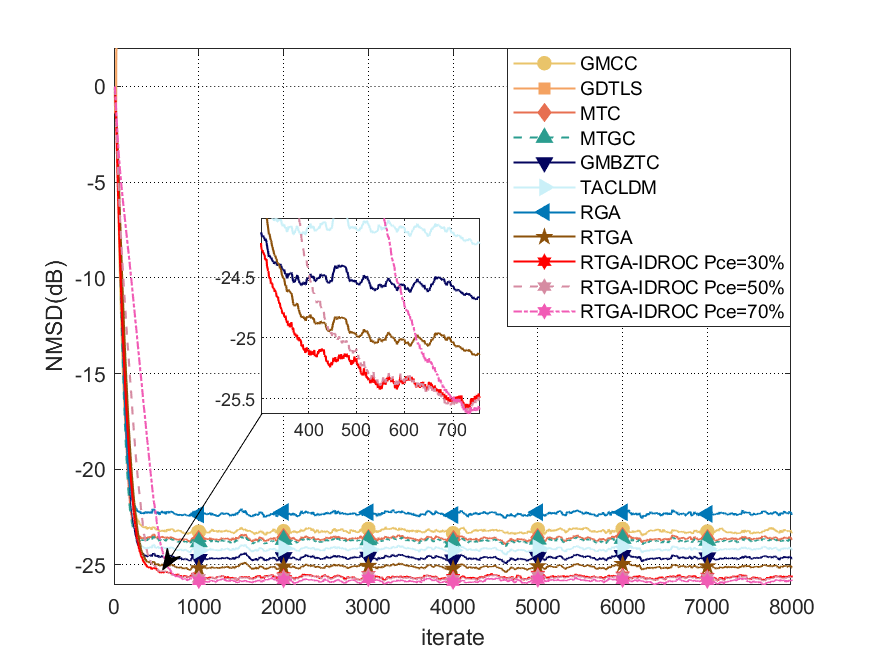}
		\\\text{(d)} 
	\end{minipage}
	\caption{Performance comparison under Cases 2 to 5: (a) Case 2, (b) Case 3, (c) Case 4, and (d) Case 5}
	\label{fig6}
\end{figure}

\subsection{Performance Evaluation in AEC}

\begin{figure}[t!]
	\centering
	
	\begin{minipage}{0.5\textwidth}
		\centering
		\includegraphics[width=0.7\linewidth]{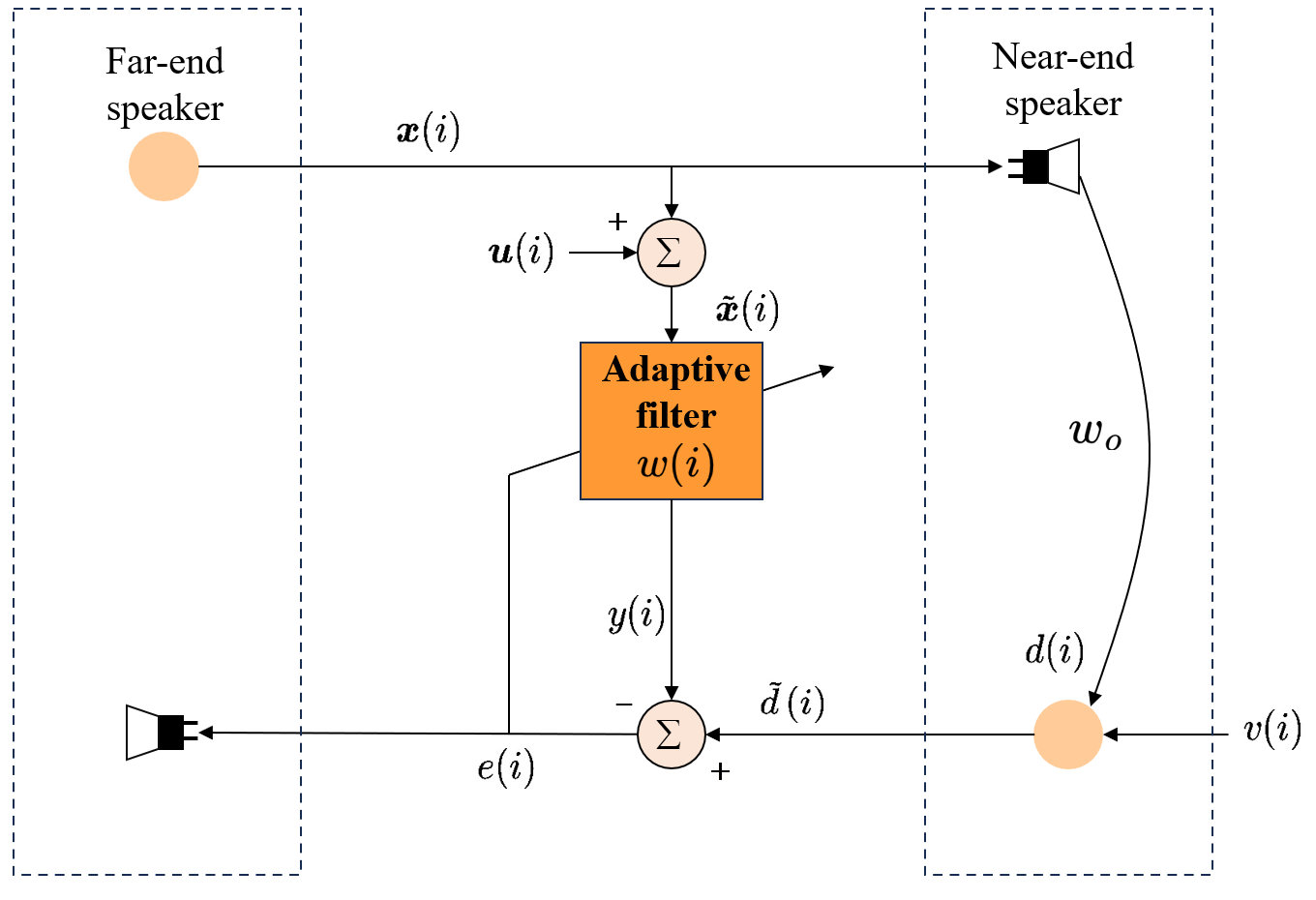}
		\\\text{(a) }
		\label{fig:sub1}
	\end{minipage}
	
	\vspace{0.5cm}
	
	\begin{minipage}{0.24\textwidth}
		\centering
		\includegraphics[width=\linewidth]{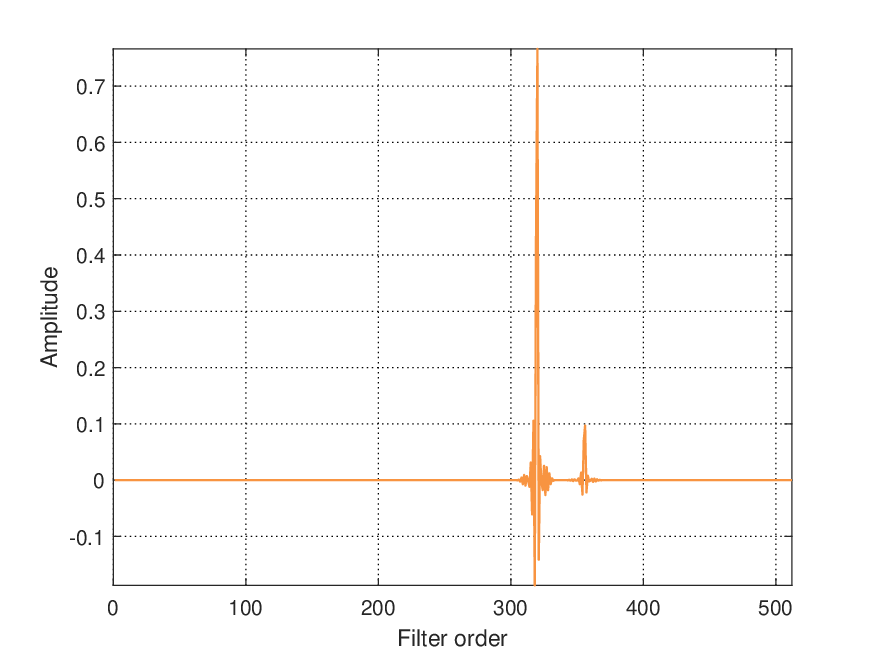}
		\\\text{(b) }
		\label{fig:sub2}
	\end{minipage}
	\hfill
	\begin{minipage}{0.24\textwidth}
		\centering
		\includegraphics[width=\linewidth]{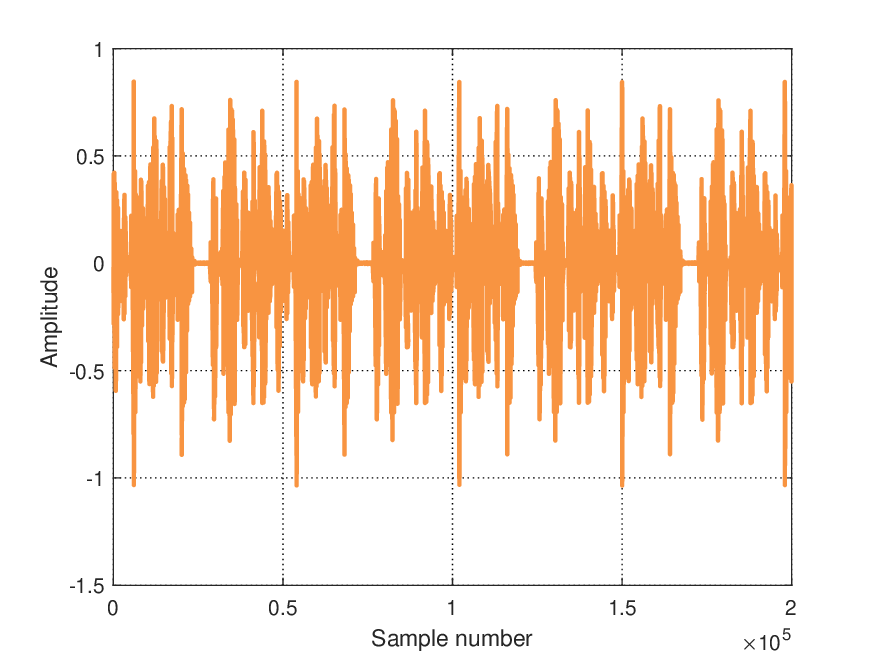}
		\\\text{(c) }
		\label{fig:sub3}
	\end{minipage}
	
	\caption{AEC environment: (a) the schematic diagram of AEC model, (b) echo path, and (c) input signal vector.}
	\label{fig7}
\end{figure}

Fig. \ref{fig7}(a) shows the schematic diagram of the acoustic echo cancellation (AEC) system. The echo path of order 512  and the real speech input signal are provided in Fig. \ref{fig7}(b) and Fig. \ref{fig7}(c), respectively.
As illustrated in Fig. \ref{fig7}(a), the input signal $\boldsymbol{x}(i)$ from the far-end speaker is corrupted by additive noise $\boldsymbol{u}(i)$ during transmission to the adaptive filter, resulting in a noisy version $\tilde{\boldsymbol{x}}(i)$ as the filter input. Note that the signal $\boldsymbol{x}(i)$ fed to the near-end speaker remains uncontaminated. This input signal $\boldsymbol{x}(i)$ is convolved with the echo path $\boldsymbol{w}_o$ to produce the ideal echo signal $d(i)$. Before reaching the adaptive filter, $d(i)$ is further impaired by noise $v(i)$, yielding the observed signal $\tilde{d}(i)$. By iteratively minimizing the error $e(i)$ between the filter output $y(i)$ and $\tilde{d}(i)$, the adaptive filter coefficients $\boldsymbol{w}(i)$ gradually converge toward $\boldsymbol{w}_o$, thereby achieving echo cancellation.

This section presents simulation results under the following configuration: 200,000 sampling points, parameters (except step size) as specified in Table II, and the input signal and echo path described previously. All results are averaged over 50 independent Monte Carlo trials.

\begin{figure}[t!]
	\centering
	\begin{minipage}{0.24\textwidth}
		\centering
		\includegraphics[width=\linewidth]{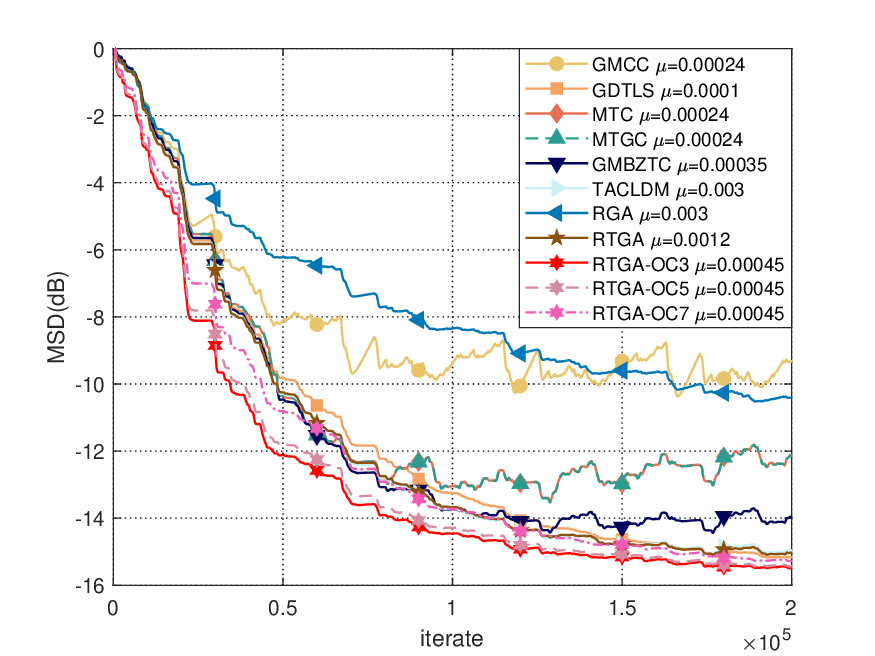}
		\\\text{(a)}
	\end{minipage}
	\hfill
	\begin{minipage}{0.24\textwidth}
		\centering
		\includegraphics[width=\linewidth]{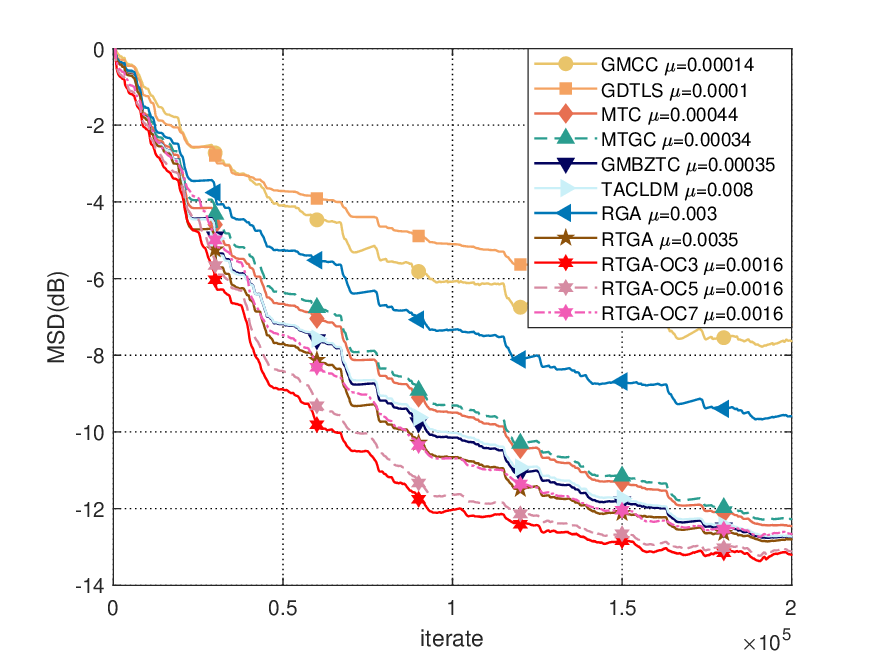}
		\\\text{(b)}
	\end{minipage}
	
	\caption{NMSD Performance comparison under: (a) Gaussian noise, (b) Laplace noise}
	\label{fig8}
\end{figure}

\subsubsection{NMSD Performance Comparison for AEC}
\mbox{}

 Figs. \ref{fig8}(a) and (b) compare the NMSD performance of the proposed algorithm against competing methods under Gaussian (Case 1) and Laplace noise (Case 3) in an AEC scenario. Consistent with earlier discussions, the IDR method reuses only the most recent 200 samples to limit memory usage. As depicted in Fig. \ref{fig8}(a), under Gaussian noise, the proposed algorithm converges noticeably faster and attains a marginally lower steady-state error compared to other methods. Specifically, using -10dB as the baseline, RTGA-IDROC(30\%), RTGA-IDROC(50\%), and RTGA-IDROC(70\%) require about 500, 700, and 1000 iterations, respectively, while other TLS-based algorithms need over 1300 iterations. These results reaffirm the efficacy of the IDR strategy and the superior performance of RTGA-IDROC. Moreover, Fig. \ref{fig8}(b) demonstrates that the proposed algorithm also maintains a clear advantage in Laplace noise, further establishing its robustness under generalized Gaussian noise conditions.

\subsubsection{ERLE Performance Comparison for AEC}
\mbox{}

\begin{figure}[t!]
	\centering
	\begin{minipage}{0.5\textwidth}
		\centering
		\includegraphics[width=\linewidth]{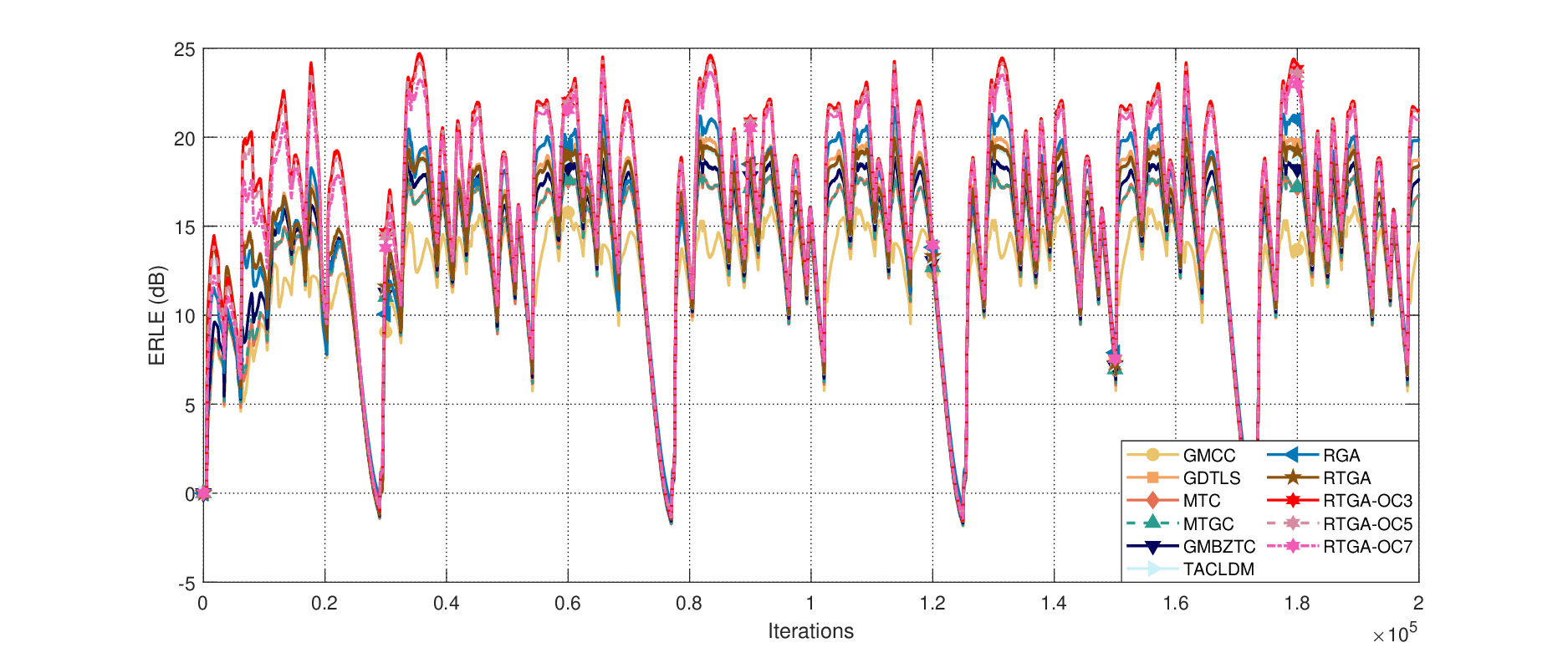}
		\\\text{(a)}
	\end{minipage}
	\hfill
	\begin{minipage}{0.5\textwidth}
		\centering
		\includegraphics[width=\linewidth]{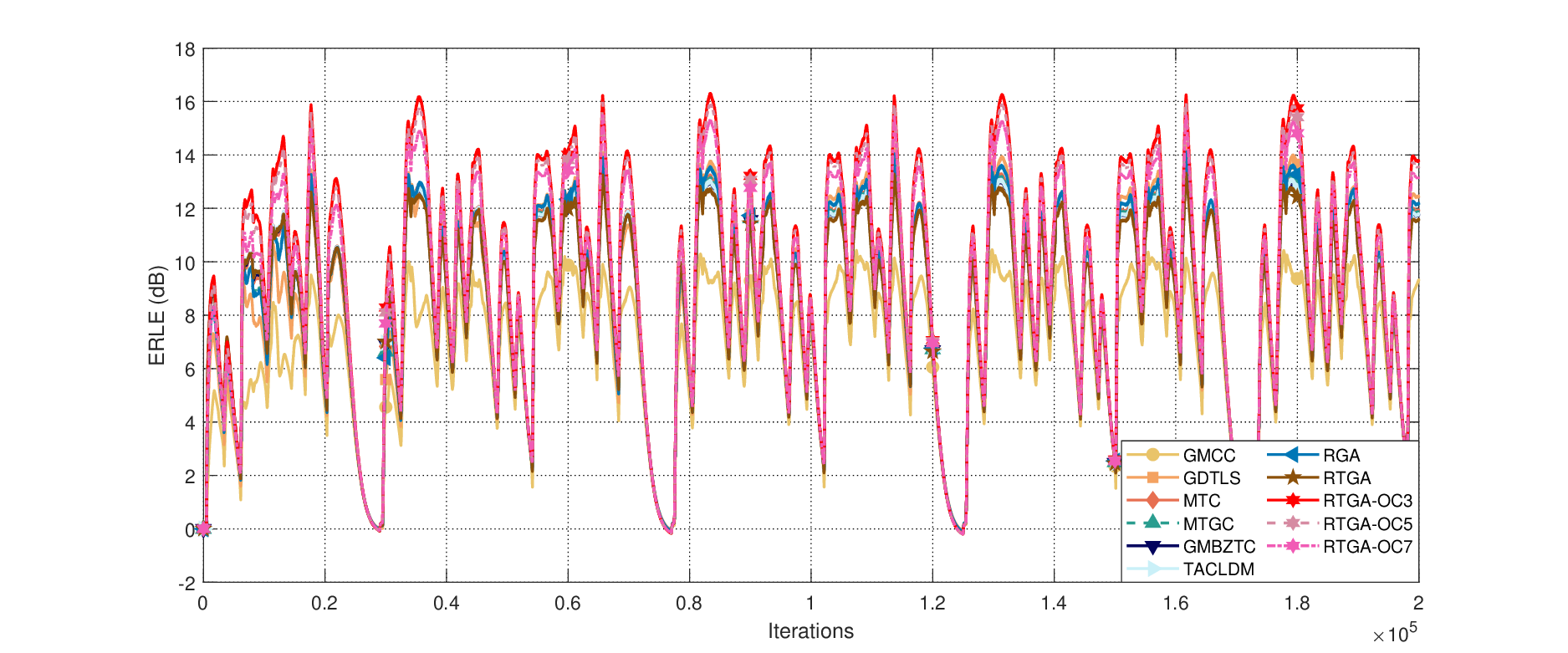}
		\\\text{(b)}
	\end{minipage}
	
	\caption{ERLE Performance comparison under: (a) Gaussian noise, (b) Laplace noise}
	\label{fig9}
\end{figure}

The echo return loss enhancement (ERLE) performance in AEC applications is investigated in this section. The ERLE is given by
\begin{equation}
	\text{ERLE}(dB) = 10 \log_{10} \left( \frac{E[d^2(i)]}{E[e^2(i)]} \right)
\end{equation}
where $E[e^2(i)]$ is obtained through the recursive relation $E[e^2(i)] = 0.999 E[e^2(i-1)] + 0.001 e^2(i)$, and $E[d^2(i)]$ is computed employing an analogous procedure. 

Figs. \ref{fig9}(a) and (b) illustrate the ERLE performance of the proposed algorithm compared with other competing algorithms in AEC applications, under background noise of Gaussian noise (Case 1) and Laplacian noise (Case 3), respectively. The parameter settings remain consistent with the previous section. As shown in Figs. \ref{fig9}(a) and (b), the proposed algorithm achieves superior ERLE performance under both Gaussian noise and generalized Gaussian noise interference, further validating its excellent performance in AEC scenarios.

\begin{figure}[t!]
	\centering
	\begin{minipage}{0.24\textwidth}
		\centering
		\includegraphics[width=\linewidth]{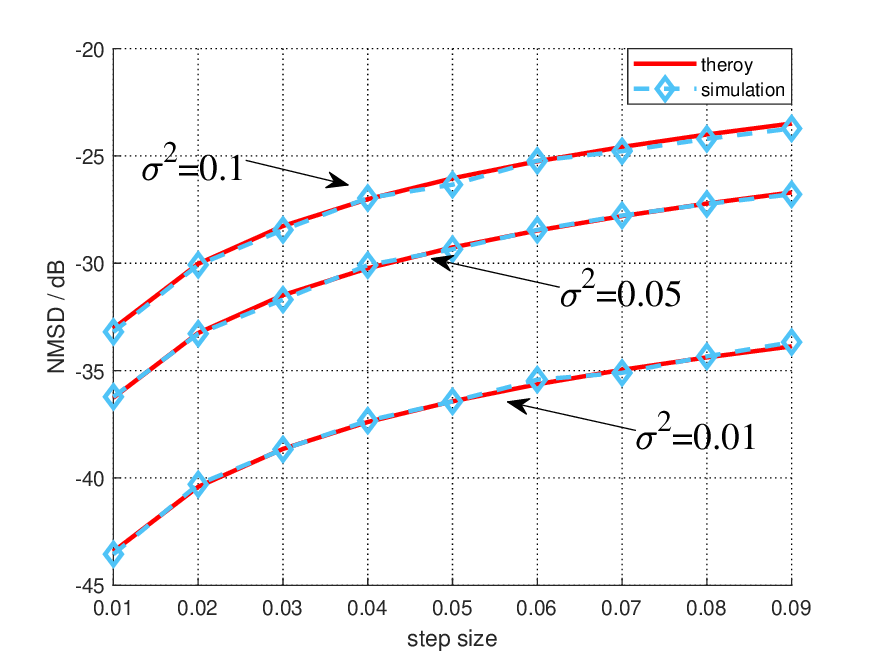}
		\\\text{(a)}
	\end{minipage}
	\hfill
	\begin{minipage}{0.24\textwidth}
		\centering
		\includegraphics[width=\linewidth]{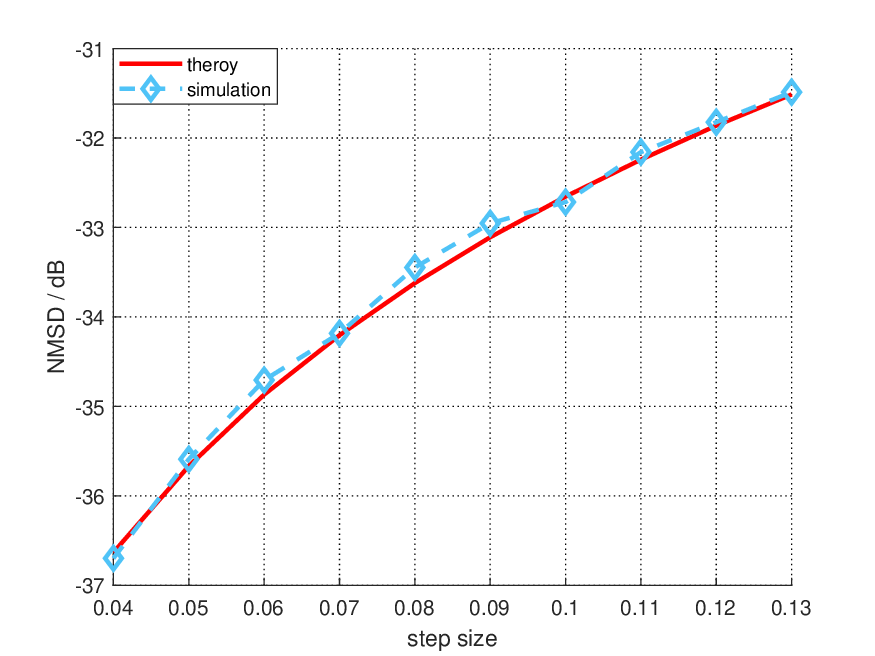}
		\\\text{(b)}
	\end{minipage}
	
	\caption{Validation of the theoretical steady-state MSD with parameter setting: (a) $a=-100, b=2, c=0.1$ and (b) $a=-100,b=1.9,c=0.1$}
	\label{fig77}
\end{figure}

\subsection{Theoretical Verification}
This section validates the consistency between the theoretical steady-state MSD of the RTGA-OC algorithm and its simulated results through computer simulations. The experiments are conducted with 200 independent MC runs and 30,000 sampling points, with specific parameter settings provided in the figure captions. Fig. \ref{fig77}(a) illustrates the close match between theoretical and simulated steady-state values under Gaussian noise with varying input and output noise variances, where $\sigma^2_i = \sigma^2_o = \sigma^2$ for each curve. To further demonstrate the validity of the theoretical analysis under generalized Gaussian noise, Fig. \ref{fig77}(b) shows the comparison under Laplace noise, where the input is Gaussian noise with variance 0.1 and the output is Laplace noise with variance 0.1. As indicated in Figs. \ref{fig77}(a) and (b), the theoretical values align closely with simulation results in both Gaussian and Laplace noise environments, confirming the correctness of the theoretical derivations.

\section{Conclusion}

In this paper, to address the issue of performance degradation in existing algorithms under the EIV model when noise environments change, we propose the RTGA-IDROC algorithm. This algorithm not only inherits the flexibility of the RGA method but also effectively mitigates estimation bias caused by input noise. Furthermore, to enhance the convergence speed in practical applications, we developed a novel data reuse (DR) method and controlled the associated increase in computational complexity using the OC strategy. Simulation results demonstrate that the RTGA-IDROC algorithm achieves significantly improved convergence speed without a substantial rise in computational complexity, thereby validating its superior performance.

\small
\bibliographystyle{IEEEtran}
\bibliography{IEEEabrv,ref}

\end{document}